\documentclass[12pt,a4paper]{article}
\usepackage[left=1in,right=1in,top=1in,bottom=1in]{geometry}
\usepackage{natbib, graphicx, color, setspace, amssymb, amsmath, amsthm, amstext, booktabs, fancyhdr, multirow, float, bm, lineno, xr,  mathtools,gensymb} 
\usepackage{graphicx} % Required for inserting images
\usepackage{enumerate} 
\usepackage{times} 
\usepackage{caption}
\usepackage{subcaption}
\usepackage{comment}
\usepackage[ruled,vlined,onelanguage]{algorithm2e}

\doublespace
\allowdisplaybreaks %Allows equations to flow across pages, and suppresses va break via \\* command
\usepackage{pdflscape}
\usepackage{placeins} %allows barriers to be placed for tighter control of figure position
\usepackage{authblk} %Allows author to have block and multiple affiliations

\newcommand*\patchAmsMathEnvironmentForLineno[1]{%
  \expandafter\let\csname old#1\expandafter\endcsname\csname #1\endcsname
  \expandafter\let\csname oldend#1\expandafter\endcsname\csname end#1\endcsname
  \renewenvironment{#1}%
     {\linenomath\csname old#1\endcsname}%
     {\csname oldend#1\endcsname\endlinenomath}}% 
\newcommand*\patchBothAmsMathEnvironmentsForLineno[1]{%
  \patchAmsMathEnvironmentForLineno{#1}%
  \patchAmsMathEnvironmentForLineno{#1*}}%
\AtBeginDocument{%
\patchBothAmsMathEnvironmentsForLineno{equation}%
\patchBothAmsMathEnvironmentsForLineno{align}%
\patchBothAmsMathEnvironmentsForLineno{flalign}%
\patchBothAmsMathEnvironmentsForLineno{alignat}%
\patchBothAmsMathEnvironmentsForLineno{gather}%
\patchBothAmsMathEnvironmentsForLineno{multline}%
}

\usepackage{hyperref}
\hypersetup{colorlinks = true, linkcolor = blue, urlcolor = blue, citecolor = blue}
\def\lsk{\left(}
\def\rsk{\right)}
\def\lbk{\left \{ }
\def\rbk{\right \} }
\def\lmk{\left [ }
\def\rmk{\right ] }
\def\lak{\left | }
\def\rak{\right | }

\newcommand{\bmbeta}{\bm{\beta}}
\newcommand{\bmepsilon}{\bm{\epsilon}}
\newcommand{\bmSigma}{\bm{\Sigma}}
\newcommand{\bmmu}{\bm{\mu}}
\newcommand{\bmtheta}{\bm{\theta}}
\newcommand{\bmrho}{\bm{\rho}}
\newcommand{\bmL}{\bm{L}}

\usepackage{xcite}
\makeatletter
\newcommand*{\addFileDependency}[1]{% argument=file name and extension
  \typeout{(#1)}% latexmk will find this if $recorder=0 (however, in that case, it will ignore #1 if it is a .aux or .pdf file etc and it exists! if it doesn't exist, it will appear in the list of dependents regardless)
  \@addtofilelist{#1}% if you want it to appear in \listfiles, not really necessary and latexmk doesn't use this
  \IfFileExists{#1}{}{\typeout{No file #1.}}% latexmk will find this message if #1 doesn't exist (yet)
}
\makeatother

\newcommand*{\myexternaldocument}[1]{%
    \externaldocument{#1}%
    \addFileDependency{#1.tex}%
    \addFileDependency{#1.aux}%
}
\myexternaldocument{supplementary}

\begin{document}
    % \title{Cokrig-and-Regress with Bootstrap for Spatially Misaligned Data}
    %\title{Cokrig-and-Regress for Spatially Misaligned Data with Multiple Covariates}
    \title{Cokrig-and-Regress for Spatially Misaligned Environmental Data}
%    \author{}
%\begin{comment}
    \author[1]{Zhi Yang Tho\thanks{Corresponding author: ZhiYang.Tho@anu.edu.au; Research School of Finance, Actuarial Studies and Statistics, The Australian National University, Acton 2601, ACT, Australia.}} 
	\author[1]{Francis K.C. Hui} 
	\author[1]{A.H. Welsh}
        \author[1]{Tao Zou}
	\affil[1]{Research School of Finance, Actuarial Studies and Statistics, The Australian National University, Canberra, Australia}
%	\affil[2]{Address 2}
%\end{comment}
	\date{}
	\maketitle
 
 \begin{abstract}
    % Understanding the relationship between the response and covariates in the analysis of spatial data is crucial in various fields. 
    Spatially misaligned data, where the response and covariates are observed at different spatial locations, commonly arise in many environmental studies.
    % such as meteorology and environmental studies. 
    Much of the statistical literature on handling spatially misaligned data has been devoted to the case of a single covariate and a linear relationship between the response and this covariate. 
    Motivated by spatially misaligned data collected on air pollution and weather in China, 
    % we study regression models for spatially misaligned data involving multiple correlated covariates. 
    we propose a cokrig-and-regress (CNR) method to estimate spatial regression models involving multiple covariates and potentially non-linear associations. The CNR estimator is constructed by replacing the unobserved covariates (at the response locations) by their cokriging predictor derived from the observed but misaligned covariates under a multivariate Gaussian assumption, where a generalized Kronecker product covariance is used to account for spatial correlations within and between covariates. 
    A parametric bootstrap approach is employed to bias-correct the CNR estimates of the spatial covariance parameters and for uncertainty quantification.
    % % these cross-correlations and 
    % the additional uncertainty from the prediction of unobserved covariates. 
    % which result in improved coverage probabilities for the regression model. 
    Simulation studies demonstrate that CNR 
    % approach combined with the parametric bootstrap 
    outperforms several existing methods for handling spatially misaligned data, such as nearest-neighbor interpolation.
    % , as well as a method that ignores the correlation between covariates.
    Applying CNR to the spatially misaligned air pollution and weather data in China reveals a number of non-linear relationships between $\text{PM}_{2.5}$ concentration and several meteorological covariates.
    % , while the confidence intervals based on the parametric bootstrap method are consistently wider than intervals constructed using the naive variance estimator.
    
\end{abstract}

\section{Introduction} \label{sec:intro}
Spatial regression models are commonly used to study the relationship between a response and one or more covariates while accounting for potential (residual) spatial correlation between the responses \citep{cressie2015statistics}, with applications in ecology, epidemiology, and environmental studies among many fields.
% through the inclusion of spatial random effects. 
%\citep{mardiaANDmarshall1984}
% Traditionally, and arguably still to this day, 
Much of the current literature on 
% methods and applications of 
spatial regression modeling assumes the response and the covariates are observed at the same spatial location. However, improving technologies means it is becoming increasingly common for the response and covariates to come from different data sources. One issue that can arise as a result of this is spatial misalignment between the response and the covariates \citep{jhunETAL2015,liuETAL2020,greenstoneETAL2022} i.e., the response and the covariates are observed at two different sets of spatial locations. As a motivating example, we consider environmental data collected on pollutant concentration (response) and meteorological variables (covariates) in China, where the pollution monitoring stations have different spatial locations to the weather stations.
% ; see Figure \ref{fig:pollution_vs_meteorological_locations} for a visualization of this. 
% Spatial misalignment though can arise in other settings such as environmental studies and epidemiology.

% Various approaches have been used to overcome spatial misalignment. 
In environmental science, the most common approach for overcoming spatial misalignment is based on nearest-neighbor interpolation, where the unobserved meteorological covariates at each pollution station are predicted to be the same as those observed at the closest meteorological station \citep[e.g.,][]{jhunETAL2015,greenstoneETAL2022}. 
%zhangETAL2017,borgeETAL2019
In related works, \citet{reichETAL2011} and \citet{liuETAL2020} predicted each meteorological covariate separately using kriging techniques to construct spatially aligned datasets. These studies do not focus on the spatial misalignment problem \emph{per-se}, and so they treat the predicted/interpolated covariates as fixed when conducting inference. Ignoring this prediction uncertainty however risks underestimating the variability of the model parameter estimates. 
%In a parallel line of research, \cite{zhuETAL2003,youngETAL2009,camelettiETAL2019} considered various ways to deal with the change of support problem where the response and covariates are given as areal data and geostatistical data, respectively. In this article though, we focus on the case where both response and covariates are geostatistical data. 
A more systematic approach to handling spatially misaligned data is the krige-and-regress (KNR) method of \citet{madsenETAL2008}, who also used kriging to spatially align the covariate to the response, but developed a Monte Carlo method to estimate the variance of the regression coefficient estimator which takes into account additional variability due to the predicted covariate. \cite{szpiro2011} further developed KNR and proposed three parametric bootstrap techniques to obtain the corrected variance estimator of the regression coefficient estimator, based on framing the uncertainty in the predicted covariate as a form of measurement error. More recently, \cite{pouliot2023} studied a bootstrap approach for KNR but under a survey sampling framework.
% , and accounted for the variability of the estimated parameters associated with the distribution of the covariate itself. 
In all of these papers on KNR, only a single misaligned covariate was considered, along with a linear response-covariate relationship. These assumptions can be too restrictive in practice though, as many environmental studies involve multiple misaligned and correlated covariates, and potentially non-linear associations between the response and the covariates; see the motivating pollution study in China in Section \ref{sec:real_data} for instance.  

In this article, to overcome the above issues we propose a new cokrig-and-regress (CNR) method for spatially misaligned data with multiple correlated covariates.
% , where both the response and covariates are geostatistical data.
% possess highly non-linear relationship with the response. The non-linear associations are captured through general vector-valued functions for the covariates used in the spatial linear mixed model. 
Building upon \cite{madsenETAL2008}, the CNR method employs a cokriging predictor \citep{steinANDcorsten1991} 
%journel1978mining,
of the unobserved covariates at the response locations.% under a multivariate Gaussian assumption.
In particular, to utilize the spatial information both within the same covariate and between covariates, we assume the joint vector of all covariates at the response and covariate locations follows a multivariate Gaussian distribution with a covariance matrix constructed based on a generalized Kronecker product form \citep{martinez2013,bonatETAL2021}. 
%bonat2016
This induces cross-correlations between covariates in such a way that it preserves the marginal covariance matrix of each covariate. In our setting of multiple spatial covariates, we use the Mat\'{e}rn class of correlation functions \citep{Matern1960} to model the marginal spatial correlation, while the cross-correlation matrix between covariates is left unstructured. 
%Additionally, the form of the generalized Kronecker product covariance matrix facilitates a two-step estimation method for the marginal mean and covariance parameters of each covariate, followed by estimating the positive-definite cross-correlation matrix. The generalized Kronecker product form is also computationally attractive as its precision matrix can be computed efficiently in likelihood computation and cokriging 
% the computational advantage of the generalized Kronecker product form in modeling the cross-covariance between two spatial covariates compared to other forms has also been noted by \citep{ribeiroETAL2022}.
The predicted covariates obtained from cokriging are used in place of the unobserved covariates in a spatial linear mixed model for the response, where we include a spatial random effect whose covariance is (again) characterized by a Mat\'{e}rn correlation function. 
% to obtain the CNR estimator of the regression coefficients associated with the potentially non-linear functions of the covariates. Therefore, the proposed method establishes a connection between the prediction and regression problems in geostatistics. 
To account for potentially non-linear functions of the covariates, we allow polynomial or spline basis functions for one or more of the covariates in the spatial linear mixed model.
% for each covariate based on the CNR estimates of the regression coefficients can be constructed to visualize the non-linear effect of the covariate on the response by varying the covariate value over a pre-specified range. 
% In both the model for the multiple covariates and the spatial linear mixed model, we make use of the Mat\'{e}rn class of correlation function \citep{Mat\'{e}rn1960} to model the spatial correlation of each covariate and the spatial random effects.
% which is first introduced by \cite{handcockANDstein1993} into the field of spatial statistics. The Mat\'{e}rn class is a wide class of correlation functions as it includes various subclasses of correlation functions as special cases, e.g., the exponential and squared exponential correlations. In contrast to the work of \cite{madsenETAL2008} that requires the choice of parametric variogram models via manual inspection of the empirical variogram, the assumption of the Mat\'{e}rn class correlation function provides a more structured way to capture the spatial correlation and facilitate the relevant parameter estimation. 
For uncertainty quantification, we follow the ideas of \citet{szpiro2011} and \citet{pouliot2023} and develop a parametric bootstrap approach to construct confidence intervals/bands for the regression coefficients/smoothing functions in CNR. 
%Unlike \cite{pouliot2023} who focused on a sample survey setting, we do not employ a sampling with replacement approach for bootstrapping. Instead, we draw joint samples of covariates at all spatial locations from a multivariate Gaussian distribution characterized by the generalized Kronecker product covariance form discussed above, and then simulate responses at the misaligned spatial locations only based on the spatial linear mixed model. 
% Our parametric bootstrap accounts for the additional uncertainty from using the prediction of unobserved covariates in the CNR method, by including an estimation step for the parameters of the covariates' joint distribution, and a prediction step based on cokriging in the bootstrap procedure. 
We also use the parametric bootstrap to bias-correct the spatial covariance parameter estimates in the spatial linear mixed model, which turns out to play an important role in improving the inferential performance of the CNR method. 
% This is because by treating the spatial misalignment problem as a form of measurement error, the estimates of the spatial covariance parameters in our setting are prone to overestimation, which is similar to the findings in the measurement error literature \citep[e.g.][]{liETAL2009,caoETAL2023}. 
%This motivates the use of a bootstrap to (also) perform bias correction. 
%In the measurement error literature, it has also been shown that positive biases can arise for estimates of the spatial covariance parameters in a spatial linear mixed model when measurement error is present in the covariates (see the theoretical analysis of \citeauthor{liETAL2009}, \citeyear{liETAL2009}, on this topic, and similar positive biases in the numerical study of \citeauthor{caoETAL2023}, \citeyear{caoETAL2023}). 

% We carry out simulation studies to assess the estimation and inferential performance of the proposed CNR estimator.
% and the alternative bootstrap estimator under the spatial misalignment problem. 

Simulation studies show CNR performs well in recovering the underlying response-covariate relationship when data are spatially misaligned, while existing approaches such as nearest-neighbor interpolation can suffer from substantial biases. The parametric bootstrap is also shown to be effective in
% in bias-correcting the spatial covariance parameter estimates, which results 
producing confidence intervals with empirical coverage probability close to the nominal level, 
% that both estimators have similar and satisfactory finite sample estimation performance. We also compare the inference performance 
% of the parametric bootstrap approach and the naive variance estimator that ignores the cross-correlation and the uncertainty of cokriging predictor, 
greatly outperforming naive variance estimators that ignore the uncertainty in correcting for the spatial misalignment, as well as a variance estimator that ignores the spatial cross-correlation between covariates. 
% in terms of empirical coverage probability of the confidence intervals in the simulation studies. 
Applying CNR to the spatially misaligned air pollution and meteorological data from China, we find evidence of complex non-linear associations between $\text{PM}_{2.5}$ concentration and meteorological covariates such as temperature and precipitation,
% by using natural cubic splines for the covariates. 
%The generalized Kronecker product form is able to capture much of the correlation pattern between the meteorological covariates uncovered during exploratory data analysis, while
% which indicates the importance of accounting for such cross-correlation in the inference. 
while the resulting spatial maps constructed from cokriging agree with scientific explanations for the spatial distribution of these meteorological covariates in China.
% and provide support for the assumed Mat\'{e}rn class correlation function. 
The confidence bands for the estimated smoothing functions of the covariates using the parametric bootstrap approach are always wider than those using the naive variance estimator, which is similar to the findings from our simulation studies where the naive variance estimator underestimates the uncertainty of the CNR estimator.
% , as the latter is likely to underestimate the true variability of the conditional smoothers based on the simulation results.

The rest of this article is organized as follows. Section \ref{sec:model} introduces the model for the spatial misalignment problem and the generalized Kronecker product covariance matrix. Section \ref{sec:CNR} develops the CNR method and the parametric bootstrap procedure. Simulation studies are given in Section \ref{sec:simulation}, while an application to spatially misaligned air pollution and meteorological data from China is provided in Section \ref{sec:real_data}. Section \ref{sec:conclusion} offers some concluding remarks. 
% Additional results for the real data example can be found in the Appendix.

\section{Data and Model Set-Up} \label{sec:model}
Let $S = \{s_1,\cdots, s_N\} \subset D$ denote a set of spatial locations in some domain of interest $D$, and suppose $\bm{y} = (y_1,\cdots,y_N)^\top = \left( y(s_1), \cdots, y(s_N) \right)^\top$ is an $N$-vector of responses observed at locations in $S$. In addition to the response, consider a set of $K$ covariates where $\bm{x}_k = (x_{1k}, \cdots, x_{Nk})^\top = (x_k(s_1), \cdots, x_k(s_N))^\top$ denotes the set of values for the $k$-th covariate at locations in $S$ for $k=1,\cdots,K$. We assume the response follows the spatial linear mixed model
% at the $i$-th location,
\begin{equation}
    y_i = \beta_0 + \sum_{k=1}^{K} \bm{f}_k(x_{ik})^\top \bmbeta_k  + \rho_i + \epsilon_i, \text{ for } i=1,\cdots,N,
    \label{eq:model_y}
\end{equation}
where $\beta_0$ is an intercept term, $\bm{f}_k(x_{ik})$ is a known vector-valued function evaluated at $x_{ik}$ for $k=1,\cdots,K$, and $\bmbeta_k$ denotes the corresponding regression coefficients. Equation \eqref{eq:model_y} includes the special case of $\bm{f}_k(x_{ik})^\top \bmbeta_k =  x_{ik} \beta_k$ and $K = 1$; such a model is the focus for much statistical research on spatially misaligned data, including the KNR method of \citet{madsenETAL2008}. In this article however, we focus on a more general setting with $K > 1$ and potentially non-linear response-covariate relationships as modeled through the vector-valued function $\bm{f}_k(x_{ik})$. For instance, in the motivating air pollution data in China, we have $K = 7$ meteorological covariates and consider second order polynomial or natural cubic spline basis functions, as part of $\bm{f}_k(x_{ik})$. 
Finally, $\bmrho = (\rho_1, \cdots, \rho_N)^\top = (\rho(s_1),\cdots, \rho(s_N))^\top$ is a vector of spatial random effects which we assume follows a multivariate Gaussian distribution with mean zero and spatial covariance matrix $\bmSigma_\rho$, while $\bmepsilon = (\epsilon_1,\cdots, \epsilon_N)^\top = (\epsilon(s_1),\cdots,\epsilon(s_N))^\top$ is a vector of independent and identically distributed (i.i.d) residuals assumed to be normally distributed with mean zero and variance $\tau_\epsilon > 0$.
% which is independent of $\bmrho$. 
The covariance matrix $\bmSigma_\rho$ captures the residual spatial correlation in the spatial linear mixed model, and we discuss its form shortly. 

In the setting of spatially misaligned data, the response and covariates are not observed at the same spatial locations. We formalize this as follows. Assume at locations in $S$ only the response vector $\bm{y}$ is observed, while the covariates $\bm{x}_k$ are missing for $k=1,\cdots,K$. Let $\tilde{S} = \{\tilde{s}_1,\cdots, \tilde{s}_M\} \subset D$ denote another set of spatial locations such that $S \cap \tilde{S} = \emptyset$, where we record $\tilde{\bm{x}}_k = (\tilde{x}_{1k}, \cdots, \tilde{x}_{Mk})^\top = (x_k(\tilde{s}_1), \cdots, x_k(\tilde{s}_M))^\top$ as covariate values for $k=1,\cdots,K$. We aim to estimate and perform inference on the coefficients $\bmbeta_k$, and functions thereof, in equation \eqref{eq:model_y}, when the responses and the covariates are spatially misaligned as described above. %That is, $\bm{y}$ and $\{\tilde{\bm{x}}_k : k = 1,\cdots,K\}$ have been observed at $S$ and $\tilde{S}$, respectively, while $\{\bm{x}_k: k = 1,\cdots,K\}$ at $S$ are unobserved. 

% If $\{\bm{x}_k: k = 1,\cdots,K\}$ has been observed, the estimation of $\bmbeta_k$ can be done by simply fitting a spatial linear mixed model to $\bm{y}$ and $\{\bm{x}_k: k = 1,\cdots,K\}$ observed at the same set locations in $S$. However, when $\{\bm{x}_k: k = 1,\cdots,K\}$ is unobserved, 
To overcome spatial misalignment, we begin by first
% will need to make use of the misaligned covariates $\{\tilde{\bm{x}}_k : k = 1,\cdots,K\}$ observed at $\tilde{S}$ by 
specifying a model for each covariate at all spatial locations i.e., $(\tilde{\bm{x}}_k^\top, \bm{x}_k^\top)^\top$. Specifically, for $k = 1,\cdots,K$, we assume $(\tilde{\bm{x}}_k^\top, \bm{x}_k^\top)^\top$ follows a multivariate Gaussian distribution with mean vector $\mu_k \bm{1}_{M+N}$ and covariance matrix 
\begin{equation*}
    \bmSigma_k = 
    \begin{pmatrix}
    \bmSigma_{k,\tilde{S}} & \bmSigma_{k,\tilde{S} S} \\
    \bmSigma_{k,S \tilde{S} } & \bmSigma_{k,S}
    \end{pmatrix} =
    \begin{pmatrix}
        \mathrm{Cov}(\tilde{\bm{x}}_k,\tilde{\bm{x}}_k) & \mathrm{Cov}(\tilde{\bm{x}}_k,\bm{x}_k) \\
        \mathrm{Cov}(\bm{x}_k,\tilde{\bm{x}}_k) & \mathrm{Cov}(\bm{x}_k,\bm{x}_k)
    \end{pmatrix},
\end{equation*}
where $\bm{1}_{M+N}$ is an $(M+N)$-vector of ones, $\mathrm{Cov}(x_k(s), x_k(s')) = \sigma_k^2 \mathcal{M}(\| s - s'\|; \nu_k, \alpha_k) + \tau_k 1_{\{s = s'\}}$ for $s,s' \in D$ and $\| s-s' \|$ denotes a distance measure between the two locations $s$ and $s'$. Here we choose $\mathcal{M}(d; \nu, \alpha) = 2^{1-\nu} (\alpha d)^{\nu} K_{\nu}(\alpha d)/ \Gamma(\nu)$ to be the Mat\'{e}rn class of correlation functions with smoothness $\nu > 0$ and range $\alpha > 0$ parameters, where $K_{\nu}(\cdot)$ is the modified Bessel function of order $\nu$, and $\Gamma(\cdot)$ is the Gamma function. The quantity $\sigma^2 > 0$ denotes the Mat\'{e}rn variance, while a nugget effect is included such that $1_{\{\cdot\}}$ denotes the indicator function and $\tau > 0$ denotes the nugget parameter. 

Returning to the spatial linear mixed model in \eqref{eq:model_y}, we also use a Mat\'{e}rn correlation function to characterize the spatial covariance matrix of the spatial random effects i.e., the elements of $\bmSigma_\rho$ are such that $\mathrm{Cov}\{\rho(s),\rho(s')\} = \sigma^2_\rho \mathcal{M}(\| s-s' \|; \nu_\rho, \alpha_\rho)$. The Mat\'{e}rn correlation function is a well-established and widely used method for characterizing the spatial correlation structure in spatial regression models \citep{INLApackage}, 
% \citep{guttorpANDgneiting2006}, %hui2022gee
with special cases being
% As the Mat\'{e}rn correlation function only depends on the distance between the two locations, it is a class of isotropic correlation function. The popularity of the Mat\'{e}rn correlation function stems from its flexibility in modeling spatial correlation and the interpretability of its parameters. With regard to the literature on Gaussian processes, the smoothness parameter controls the smoothness of the underlying spatial process by directly governing the correlation at small distances, while the range parameter determines the decay rate of the correlation at large distances. More specifically, if a spatial process on a Euclidean space has a Mat\'{e}rn correlation function with smoothness parameter $\nu$, then it is $\ceil{\nu} - 1$ times mean square differentiable, where $\ceil{\nu}$ denotes the largest integer less than or equal to $\nu$. 
% The flexibility of the Mat\'{e}rn class of correlation functions is also evident through its 
% inclusion of a wide range of subclasses of correlation functions as special cases. The 
the exponential correlation function when $\nu = 1/2$,
% i.e., $\mathcal{M}(d;1/2, \alpha) = \exp(-\alpha d)$
and the squared exponential or Gaussian correlation function when $\nu \to \infty$.
% i.e., $\mathcal{M}(d;\infty, \alpha) \to \exp\{ -(\alpha d)^2 \}$. 
%The case of $\nu = 1$ has also become popular for two-dimensional spatial data with an Euclidean distance metric \citep{INLApackage}.
In practice it is quite common to fix the spatial smoothness parameter $\nu$ while estimating the variance $\sigma^2$ and spatial range $\alpha$ \citep[e.g.,][who fixed $\nu = 1$ for two-dimensional spatial data with an Euclidean distance metric]{INLApackage,bakka2018spatial}, and we adopt this approach for our spatial misalignment application while choosing $\nu$ based on exploratory data analysis. 
%For instance, fixing $\nu = 1$ has become popular for two-dimensional spatial data with an Euclidean distance metric \citep[e.g.,][]{INLApackage}.
% and \citet{bakka2018spatial} and references therein,
%\citet{lindgren2011explicit} and
%the case of . 
% \cite[see also][who demonstrated the challenges in estimating the parameters in the Mat\'{e}rn correlation function]{zhang2004}. 
% and proved that the quantity $\sigma^2 \alpha^{2\nu}$ which is important for interpolation can be estimated consistently when $\nu$ is known.

\subsection{Generalized Kronecker Product Covariance Matrix} \label{sec:generalized_kronecker}

So far, we have specified a model for spatial correlation within a covariate, which largely follows from the work of \citet{madsenETAL2008}, \citet{szpiro2011} and \citet{pouliot2023} for spatial misalignment. However, as discussed in Section \ref{sec:intro}, when $K > 1$ it is of interest to allow for the covariates to be correlated with each other. For example, we find evidence of moderate correlations between several of the meteorological covariates in our motivating air pollution data in Section \ref{sec:real_data}. Accounting for such spatial cross-correlations should improve uncertainty quantification and hence inference on the $\bmbeta_k$'s in equation \eqref{eq:model_y}.

Let $\bm{x} = ( \tilde{\bm{x}}_1^\top, \bm{x}_1^\top, \cdots, \tilde{\bm{x}}_K^\top, \bm{x}_K^\top)^\top$ denote the stacked $K(M+N)$-vector of $K$ covariates at locations in both $\tilde{S}$ and $S$, and $\bm{x}_{\tilde{S}} = (\tilde{\bm{x}}_1^\top, \cdots, \tilde{\bm{x}}_K^\top)^\top$ and $\bm{x}_{S} = (\bm{x}_1^\top, \cdots, \bm{x}_K^\top)^\top$ be the $KM$- and $KN$-vectors of covariates at locations in $\tilde{S}$ and $S$, respectively. 
Then we assume $\bm{x}$ follows a multivariate Gaussian distribution with mean vector $\bmmu \otimes \bm{1}_{M+N}$  
% , $\otimes$ is the Kronecker product operator, 
and covariance matrix 
\begin{equation}
    \bmSigma = \mathrm{Bdiag}(\bmL_1,\cdots, \bmL_K) (\bm{R} \otimes \bm{I}_{M+N}) \mathrm{Bdiag}(\bmL_1,\cdots, \bmL_K)^\top,
    \label{eq:generalized_kronecker}
\end{equation}
where $\bmmu = (\mu_1,\cdots,\mu_K)^\top$, $\bm{R}$ is a $K \times K$ correlation matrix capturing the cross-correlation among the $K$ covariates, $\bm{I}_{M+N}$ is the $(M+N) \times (M+N)$ identity matrix, $\bmL_k$ is the lower Cholesky factor of $\bmSigma_k$, and $\mathrm{Bdiag}(\cdot)$ denotes the block diagonal matrix operator. 

When $K = 1$, it is easy to see that $\bm{\Sigma}$ in equation \eqref{eq:generalized_kronecker} reduces to a single Mat\'{e}rn covariance matrix $\bm{\Sigma}_k$ discussed in the previous section. For $K > 1$, 
% with the $k$-th block being $\bmSigma_k$. 
the covariance matrix in \eqref{eq:generalized_kronecker} has the generalized Kronecker product form proposed by \citet{martinez2013}, and subsequently used by \citet{bonat2016,bonatETAL2021} among others for multivariate covariance generalized linear models. This form induces a relatively simple structure for the  
% Equation \eqref{eq:generalized_kronecker} is not only consistent with the previous assumption on the marginal covariance matrix; that is, $\mathrm{Cov}((\tilde{\bm{x}}_k^\top, \bm{x}_k^\top)^\top, (\tilde{\bm{x}}_k^\top, \bm{x}_k^\top)^\top) = \bmSigma_k$, but also 
spatial cross-correlation between two covariates: $\mathrm{Cov}\{(\tilde{\bm{x}}_{k_1}^\top, \bm{x}_{k_1}^\top)^\top, (\tilde{\bm{x}}_{k_2}^\top, \bm{x}_{k_2}^\top)^\top\} = r_{k_1k_2} \bmL_{k_1} \bmL_{k_2}^\top$ for $k_1 \neq k_2$, where $r_{k_1k_2}$ is the $(k_1,k_2)$-th element of $\bm{R}$. Although such a simple form may be less flexible than other options for modeling spatial cross-covariances \citep[see][and references therein]{salvana2020nonstationary}, 
%10.1214/14-STS487
the generalized Kronecker product does offer some notable advantages in terms of computation as well as the relatively smaller number of parameters needing to be estimated, as we shall see shortly.
% \citep[see also][for a discussion on its computational advantage when $K = 2$]{ribeiroETAL2022}. We discuss these shortly.

The assumed model for the full vector $\bm{x}$ in \eqref{eq:generalized_kronecker} implies that the vector of only the observed covariates, $\bm{x}_{\tilde{S}}$, is multivariate Gaussian distributed with mean $\bmmu \otimes \bm{1}_M$ and covariance matrix
\begin{equation}
    \bmSigma_{\tilde{S}} = \mathrm{Bdiag}(\bmL_{1,\tilde{S}},\cdots, \bmL_{K,\tilde{S}}) (\bm{R} \otimes \bm{I}_{M}) \mathrm{Bdiag}(\bmL_{1,\tilde{S}},\cdots, \bmL_{K,\tilde{S}})^\top.
    \label{eq:generalized_kronecker_tilde}
\end{equation}
This is analogous to the form of $\bmSigma$ in \eqref{eq:generalized_kronecker}, where $\bmL_{k,\tilde{S}}$ is the lower Cholesky factor of $\bmSigma_{k,\tilde{S}}$. 

Let $\bmtheta_x = (\bmmu^\top,\sigma_1^2,\cdots,\sigma_K^2, \nu_1,\cdots, \nu_K, \alpha_1,\cdots,\alpha_K, \tau_1,\cdots,\tau_K, \mathrm{vech}(\bm{R})^\top )^\top$ be the $(5K + K(K+1)/2)$-vector of parameters characterizing the joint distribution of the covariates.
% where $\mathrm{tril}(\bm{R})$ denotes the vector consisting of the lower triangular elements of $\bm{R}$. 
The generalized Kronecker product form for $\bmSigma_{\tilde{S}}$ means that the parameters $\bmtheta_x$ can be estimated directly from the marginal distribution of just the observed covariate vector $\bm{x}_{\tilde{S}}$.
%The preservation of the form of a generalized Kronecker product for $\bmSigma_{\tilde{S}}$ is beneficial for the estimation of $\bmtheta_x$, since it means that the parameters $\bmtheta_x$ can be estimated directly from the marginal distribution of just observed covariate vector $\bm{x}_{\tilde{S}}$, as opposed to, say, having to work with the full vector $\bm{x}$ and imputing the missing $\bm{x}_S$ along the way.
% using an Expectation-Maximization algorithm \citep{dempster1977maximum}. 
Additionally, the precision matrix can be computed efficiently as
$ \bmSigma_{\tilde{S}}^{-1} = \mathrm{Bdiag}(\bmL_{1,\tilde{S}}^{-1},\cdots, \bmL_{K,\tilde{S}}^{-1})^\top (\bm{R}^{-1} \otimes \bm{I}_{M}) \mathrm{Bdiag}(\bmL_{1,\tilde{S}}^{-1},\cdots, \bmL_{K,\tilde{S}}^{-1})$. 
% which we require as part of the multivariate Gaussian likelihood as well as ultimately for the proposed cokriging procedure of Section \ref{sec:CNR}. 
This only requires inverting matrices of dimensions $M \times M$ and $K \times K$, as opposed to directly inverting a matrix of dimension $MK \times MK$. 
% We will make use of this result in the following section. 
% this is useful when $M$ and $K$ are not small as in our air pollution application in Section \ref{sec:real_data}. 

\section{Cokrig-and-Regress} \label{sec:CNR}
% To address the spatial misalignment problem with more than one covariate, 
We propose cokrig-and-regress (CNR) for estimating $\bmbeta = (\beta_0,\bmbeta_1^\top, \cdots, \bmbeta_K^\top)^\top$ in equation \eqref{eq:model_y} when only $\bm{y}$ and the misaligned $\bm{x}_{\tilde{S}}$ are observed. The proposed approach can be viewed as a generalization of KNR in \cite{madsenETAL2008} to allow for $K>1$ covariates and potentially non-linear associations between the response and the covariates, while taking into account potential spatial cross-correlations between covariates.
% Based on the assumed multivariate Gaussian model for $\bm{x}$, the misaligned $\bm{x}_{\tilde{S}}$ can be used to predict the unobserved $\bm{x}_{S}$ by cokriging, since the covariates are not only spatially correlated but also cross-correlated with each other. Such a prediction would require the knowledge of unknown parameters in $\bmtheta_x$, which can be estimated based on the observed $\bm{x}_{\tilde{S}}$. Then, the predicted $\bm{x}_{S}$ can be used together with $\bm{y}$ in the fitting of model \eqref{eq:model_y} to estimate $\bmbeta$ and $\bmtheta_\epsilon = (\sigma^2_\epsilon, \nu_\epsilon, \alpha_\epsilon, \tau_\epsilon)^\top$. 

The CNR method consists of the following three steps. First, we estimate the parameters characterizing the distribution of the covariates i.e., $\bm{\theta}_x$. %In particular, the generalized Kronecker product form in \eqref{eq:generalized_kronecker_tilde} facilitates a two-step estimation approach as follows. 
Let $\bmtheta_{x,k} = (\mu_k, \sigma^2_k, \nu_k, \alpha_k, \tau_k)^\top$ be the parameters associated with the marginal distribution of the $k$-th covariate for $k=1,\cdots,K$. Then we begin by maximizing the marginal log-likelihood of $\tilde{\bm{x}}_k$ for $k=1,\cdots,K$,
\begin{align}
    \hat{\bmtheta}_{x,k} 
    % &= (\hat{\mu}_k,\hat{\sigma}^2_k, \hat{\nu}_k, \hat{\alpha}_k, \hat{\tau}_k)^\top \notag \\
    % &= \arg\max_{\bmtheta_{x,k} } l_k(\tilde{\bm{x}}_k;\bmtheta_{x,k}) \notag  \\
    &= \arg\max_{\bmtheta_{x,k} } \lbk -\frac{1}{2} \log\det(\bmSigma_{k,\tilde{S}}) - \frac{1}{2} (\tilde{\bm{x}}_k - \mu_k \bm{1}_M)^\top \bmSigma_{k,\tilde{S}}^{-1} (\tilde{\bm{x}}_k - \mu_k \bm{1}_M) \rbk,
    \label{eq:cnr_step1_theta_x}
\end{align}
where $\bmSigma_{k,\tilde{S}}$ is parameterized by $(\sigma^2_k, \nu_k, \alpha_k, \tau_k)^\top$ and constants in the log-likelihood with respect to the parameters are omitted. 
% Equivalently, we maximize the log-likelihood of the joint vector $\bm{x}_{\tilde{S}}$ by assuming $\bm{R} = \bm{I}_K$ in \eqref{eq:generalized_kronecker_tilde}. 
In Section \ref{sec:appendix_unbiased_estimating} of the supplementary material, we show that this approach results in asymptotically unbiased estimates of $\hat{\bmtheta}_{x,k}$ when the covariates are truly cross-correlated.
% i.e., $\bm{R} \neq \bm{I}_K$.
Recalling that $\bmSigma_{k,\tilde{S}}$ takes the form of a Mat\'{e}rn covariance matrix for $k=1,\cdots,K$, equation \eqref{eq:cnr_step1_theta_x} can be straightforwardly solved e.g., using the \texttt{R} package \texttt{spaMM} \citep{spaMMpackage}, where $\nu_k$ is assumed to be chosen based on exploratory data analysis and then fixed throughout.
% as discussed in Section \ref{sec:model}. 
% Note we can parallelize the estimation of $\hat{\bmtheta}_{x,k}$  across $k=1,\cdots,K$ to improve the efficiency of our CNR method. 
Next, let $\hat{\bmSigma}_{k,\tilde{S}}$ denote the estimated marginal covariance matrix of $\tilde{\bm{x}}_k$ evaluated at $\hat{\bmtheta}_{x,k}$, and $\hat{\bmL}_{k,\tilde{S}}$ be the corresponding Cholesky factor for $k=1,\cdots,K$. Writing $\hat{\bmSigma}_{\tilde{S}}(\bm{R}) = \mathrm{Bdiag}(\hat{\bmL}_{1,\tilde{S}},\cdots, \hat{\bmL}_{K,\tilde{S}}) (\bm{R} \otimes \bm{I}_{M}) \mathrm{Bdiag}(\hat{\bmL}_{1,\tilde{S}},\cdots, \hat{\bmL}_{K,\tilde{S}})^\top$ and given $\hat{\bmtheta}_{x,k}$ for $k=1,\cdots,K$,
% and $\hat{\bmmu} = (\hat{\mu}_1,\cdots,\hat{\mu}_K)^\top$ . 
then we can construct a one-step estimator for $\bm{R}$ as
% by maximizing the joint log-likelihood $l(\bm{x}_{\tilde{S}}; \bmtheta_x )$ of $\bm{x}_{\tilde{S}}$  as
\begin{align}
    \hat{\bm{R}} 
    % &= \arg \max_{\bm{R}} l(\bm{x}_{\tilde{S}}; \hat{\bmtheta}_{x,1}, \cdots, \hat{\bmtheta}_{x,K}, \bm{R} )  \notag \\
    &= \arg \max_{\bm{R}} \lbk -\frac{1}{2} \log \lak \hat{\bmSigma}_{\tilde{S}}(\bm{R}) \rak  -\frac{1}{2} (\bm{x}_{\tilde{S}} - \hat{\bmmu} \otimes \bm{1}_M)^\top \hat{\bmSigma}_{\tilde{S}}(\bm{R})^{-1} (\bm{x}_{\tilde{S}} - \hat{\bmmu} \otimes \bm{1}_M) \rbk  \notag \\ 
    &= \frac{1}{M} 
    \begin{pmatrix}
    (\tilde{\bm{x}}_1^\top - \hat{\mu}_1 \bm{1}_M^\top) \hat{\bmL}_{1,\tilde{S}}^{-1\top} \\
    \vdots \\
    (\tilde{\bm{x}}_K^\top - \hat{\mu}_K \bm{1}_M^\top) \hat{\bmL}_{K,\tilde{S}}^{-1\top} 
    \end{pmatrix}
    \begin{pmatrix}
        \hat{\bmL}_{1,\tilde{S}}^{-1}(\tilde{\bm{x}}_1 - \hat{\mu}_1 \bm{1}_M) , 
        \cdots, 
        \hat{\bmL}_{K,\tilde{S}}^{-1}(\tilde{\bm{x}}_K - \hat{\mu}_K \bm{1}_M)
    \end{pmatrix},
    \label{eq:cnr_step1_R}
\end{align}
where $\hat{\bm{\mu}} = (\hat{\mu}_1,\cdots,\hat{\mu}_K)^\top$ is the vector of estimated marginal mean parameters from \eqref{eq:cnr_step1_theta_x}; see  Section \ref{sec:appendix_unbiased_estimating} of the supplementary material for further discussion on the one-step estimator.
%By construction, $\hat{\bm{R}}$ is always a positive-definite matrix. Although the estimator in \eqref{eq:cnr_step1_R} is not itself a correlation matrix,
% i.e., its diagonal elements and off-diagonal elements are not guaranteed to be 1 and between -1 to 1 respectively, 
% the estimator is found to work well empirically in later sections. Specifically, the $\hat{\bm{R}}$ estimates obtained 
%in both the simulation studies and real data application we always found that $\hat{\bm{R}}$ was very close to a correlation matrix. Moreover, 
% that close-to-one diagonal elements and off-diagonal elements that are between -1 and 1. If required, 
%we found further standardizing $\hat{\bm{R}}$ to be a proper correlation matrix makes very little difference in our final estimation and inference results for the $\bm{\beta}_k$'s. As such, 
% one could also further standardize the estimator via
% $
%     \{ \mathrm{diag}(\hat{\bm{R}}) \}^{-1/2} \hat{\bm{R}} \{ \mathrm{diag}(\hat{\bm{R}}) \}^{-1/2},
% $
% where $\mathrm{diag}(\hat{\bm{R}})$ is a diagonal matrix consisting of the diagonal elements of $\hat{\bm{R}}$, to obtain an exact correlation matrix estimator, while we choose to work with the $\hat{\bm{R}}$ 
%in this article we work with $\hat{\bm{R}}$ as the estimator of the between-covariate correlation matrix. 
It is also worth noting that the aforementioned two-step estimation procedure for $\bm{\theta}_x$ is facilitated by the generalized Kronecker product form in \eqref{eq:generalized_kronecker_tilde}.

In the second step of CNR, we predict the unobserved $\bm{x}_{S}$ i.e., at the spatial locations where only the response was recorded, based on the observed $\bm{x}_{\tilde{S}}$ and the estimates from the first step (denoted here as $\hat{\bmtheta}_x$).
% parameters $\hat{\bmtheta}_x = (\hat{\bmmu}, \hat{\sigma}^2_1,\cdots, \hat{\sigma}^2_K, \hat{\nu}_1,\cdots, \hat{\nu}_K, \hat{\alpha}_1,\cdots, \hat{\alpha}_K, \hat{\tau}_1,\cdots,\hat{\tau}_K, \mathrm{tril}(\hat{\bm{R}})^\top)^\top$ from the above step. 
Given the assumption of 
% Recalling that $\bm{x}$ whose components are made up of the elements of $\bm{x}_{S}$ and $\bm{x}_{\tilde{S}}$ is assumed to be a 
multivariate normality, it is straightforward to obtain the prediction of $\bm{x}_{S}$ by cokriging i.e., the empirical best linear unbiased prediction of $\bm{x}_{S}$ given $\bm{x}_{\tilde{S}}$ which makes use of the spatial correlations both within and between covariates,
% In the case of the assumed multivariate Gaussian distribution, the prediction $\hat{\bm{x}}_{S}$ is given as
\begin{equation}
    \hat{\bm{x}}_{S} = \mathrm{E}(\bm{x}_S | \bm{x}_{\tilde{S}}; \hat{\bmtheta}_x)
    =\hat{\bmmu} \otimes \bm{1}_N + \hat{\bmSigma}_{S\tilde{S}} \hat{\bmSigma}_{\tilde{S}}^{-1} \lsk \bm{x}_{\tilde{S}} - \hat{\bmmu} \otimes \bm{1}_M \rsk,
    \label{eq:cnr_step2_cokriging}
\end{equation}
where $\hat{\bmSigma}_{S\tilde{S}}$ and $\hat{\bmSigma}_{\tilde{S}}$ are the estimated cross-covariance matrix between $\bm{x}_{S}$ and $\bm{x}_{\tilde{S}}$ and the estimated covariance matrix of $\bm{x}_{\tilde{S}}$, respectively, based on extracting the relevant rows and columns of $\hat{\bmSigma}$ i.e., the estimate of $\bm{\Sigma}$ in equation \eqref{eq:generalized_kronecker} evaluated at $\hat{\bmtheta}_x$ and $\hat{\bm{R}}$.
% corresponding to the elements of $\bm{x}_{S}$ and $\bm{x}_{\tilde{S}}$, respectively.
% , and $\hat{\bmSigma}$ is the estimated covariance matrix of $\bm{x}$ by replacing $\bmtheta_x \backslash \bmmu$ in $\bmSigma$ with the estimated $\hat{\bmtheta}_x \backslash \hat{\bmmu}$ from the first step. 
Again, with the generalized Kronecker product form this can be efficiently computed as we can straightforwardly show that the joint prediction of $K$ covariates based on cokriging using \eqref{eq:cnr_step2_cokriging} is equivalent to computing the predictors $\hat{\bm{x}}_{S} = (\hat{\bm{x}}_1^\top,\cdots, \hat{\bm{x}}_K^\top)^\top$ with $\hat{\bm{x}}_k = (\hat{x}_{1k},\cdots, \hat{x}_{Nk})^\top = \hat{\mu}_k \bm{1}_N + \hat{\bmSigma}_{k,S\tilde{S}} \hat{\bmSigma}_{k,\tilde{S}}^{-1} (\tilde{\bm{x}}_k - \hat{\mu}_k \bm{1}_M)$ where $\hat{\bmSigma}_{k,S\tilde{S}}$ denotes $\bmSigma_{k,S\tilde{S}}$ evaluated at $\hat{\bm{\theta}}_{x,k}$ for $k=1,\cdots,K$. 
% In other words, equation \eqref{eq:cnr_step2_cokriging} can be computed by performing kriging separately for each of the $K$ covariates. 
This equivalence can be used to reduce computational effort as we only have to deal with matrices of dimension $M \times M$ and $M \times N$ instead of $MK \times MK$ and $MK \times NK$ when performing cokriging. It is worth highlighting that such equivalence does not undermine the use of \eqref{eq:generalized_kronecker} for modeling the correlation between the covariates: explicitly taking into account the spatial cross-correlations between covariates improves our uncertainty quantification when we develop our bootstrap approach later on in Section \ref{sec:bootstrap} for inference.
% when the covariates are truly correlated.

In the final step of CNR, we 
% estimate $\hat{\bmbeta} = (\hat{\beta}_0,\hat{\bmbeta}_1^\top,\cdots, \hat{\bmbeta}_K^\top)^\top$, $\hat{\bmtheta}_\rho = (\hat{\sigma}^2_\rho, \hat{\nu}_\rho, \hat{\alpha}_\rho)^\top$, and $\hat{\tau}_\epsilon$ by 
substitute the predicted $\hat{x}_{ik}$'s into equation \eqref{eq:model_y} and maximize the log-likelihood of the spatial linear mixed model. Letting
\begin{equation*}
    \hat{\bm{B}} = 
    \begin{pmatrix}
    1 & \bm{f}_1(\hat{x}_{11})^\top & \cdots &   \bm{f}_K(\hat{x}_{1K})^\top \\
    \vdots & \vdots & \vdots & \vdots \\
    1 & \bm{f}_1(\hat{x}_{n1})^\top & \cdots &   \bm{f}_K(\hat{x}_{nK})^\top \\
    \end{pmatrix},
\end{equation*}
then we compute the estimates
\begin{align}
\hspace{-0.5cm}    (\hat{\bmbeta}^\top, \hat{\bmtheta}_\rho^\top, \hat{\tau}_\epsilon)^\top 
    % &= \arg\max_{(\bmbeta, \bmtheta_\epsilon)} l_y(\bm{y}, \hat{\bm{x}}_{S}; \bmbeta, \bmtheta_\epsilon) \notag \\
    &= \arg\max_{(\bmbeta, \bmtheta_\rho, \tau_\epsilon)} \lbk -\frac{1}{2} \log \lak \bmSigma_\rho + \tau_\epsilon \bm{I}_N \rak - \frac{1}{2} (\bm{y} - \hat{\bm{B}} \bmbeta )^\top (\bmSigma_\rho + \tau_\epsilon \bm{I}_N)^{-1} (\bm{y} - \hat{\bm{B}} \bmbeta ) \rbk, 
    % - \frac{N}{2} \log(2\pi),
    \label{eq:cnr_step3}
\end{align}
where $\bmSigma_\rho$ is parameterized by $\bmtheta_\rho = (\sigma^2_\rho, \nu_\rho, \alpha_\rho)^\top$. The regression coefficient estimators take the form of a generalized least squares estimator $\hat{\bmbeta} = \{\hat{\bm{B}}^\top (\hat{\bmSigma}_\rho + \hat{\tau}_\epsilon \bm{I}_N)^{-1} \hat{\bm{B}}\}^{-1} \hat{\bm{B}}^\top (\hat{\bmSigma}_\rho + \hat{\tau}_\epsilon \bm{I}_N)^{-1} \bm{y}$, where $\hat{\bmSigma}_\rho$ denotes the estimated covariance matrix evaluated at $\hat{\bmtheta}_\rho$. Since $\bmSigma_\rho$ has the form of a Mat\'{e}rn covariance matrix, then we adopt a similar approach to the estimation of $\hat{\bmtheta}_{x,k}$'s in \eqref{eq:cnr_step1_theta_x} for estimating the parameters in \eqref{eq:cnr_step3}.
% of the first step to solve \eqref{eq:cnr_step3}. 
% We also consider fixed $\nu_{\rho}$ chosen from exploratory analysis as discussed in Section \ref{sec:model}.
% by employing the \texttt{R} package \texttt{spaMM} and assuming $\nu_\rho$ is known where $\nu_\rho$ is selected according to exploratory data analysis.

As a final remark, when $\bm{f}_k(\cdot)^\top \bmbeta_k$ is used to model a non-linear relationship between the response and the $k$-th covariate, then for interpretation purposes we focus on the estimation and uncertainty quantification of the conditional smoothing function $\hat{\beta}_0 + \bm{f}_k(x_k)^\top \hat{\bmbeta}_k  + \sum_{l \ne k } \bm{f}_l(c_l)^\top \hat{\bmbeta}_l $, where we vary the value of $x_k$ while conditioning on the other covariates e.g., setting the $c_l$'s to their estimated mean values $\hat{\mu}_l$.

% The proposed CNR method can take advantage of available \texttt{R} package \texttt{spaMM} \citep{spaMMpackage} to perform the maximum likelihood estimation in \eqref{eq:cnr_step1_theta_x} and \eqref{eq:cnr_step3}. 
% It is also worth noting that the CNR method provides a more systematic way of estimating $\bmSigma$ and $\bmSigma_\epsilon$ through the use of the flexible Mat\'{e}rn class function compared to the KNR method of \cite{madsenETAL2008} which involves choosing appropriate parametric variogram models through manual inspection of the empirical variogram plots.

\subsection{Bias Correction and Uncertainty Quantification} \label{sec:bootstrap}
By using $\hat{x}_{ik}$'s to replace $x_{ik}$'s in the third step of CNR, the estimators for the spatial covariance parameters $\hat{\bmtheta}_\rho$ and $\hat{\tau}_\epsilon$ can be positively biased as the predicted $\hat{x}_{ik}$'s can be thought as the true $x_{ik}$'s contaminated with measurement error. This is analogous to results seen in the measurement error literature \citep[e.g.,][]{liETAL2009,caoETAL2023},
% (see the theoretical analysis of \citeauthor{liETAL2009}, \citeyear{liETAL2009} and the numerical study of \citeauthor{caoETAL2023}, \citeyear{caoETAL2023}) 
where positive biases are found for the spatial covariance parameter estimators of the spatial linear mixed model when there exists measurement error in the covariates; see also our simulation results in Section \ref{sec:simresults} where this positive bias results in overly wide confidence intervals for $\bm{\beta}$.
% As noted by \cite{szpiro2011}, the measurement error in the case of spatial misalignment is distinct from the classical independent measurement error, which hinders the use of classical measurement error correction techniques. Instead, 
To overcome this issue, we propose a parametric bootstrap to obtain bias-corrected CNR estimates for the spatial covariance parameters in the spatial linear mixed model \eqref{eq:model_y}. We discuss the details of this method shortly, but first we motivate the use of a resampling approach specific for uncertainty quantification.

To facilitate statistical inference on the relationship between the response and the covariates, we need to estimate the variance of the CNR estimators for the regression coefficients. Based on the form of the generalized least squares estimator $\hat{\bmbeta}$, a naive variance estimator can be obtained as $\hat{\bm{V}}_{\mathrm{naive}} = \{\hat{\bm{B}}^\top (\hat{\bmSigma}_\rho + \hat{\tau}_\epsilon \bm{I}_N)^{-1} \hat{\bm{B}}\}^{-1}$. However, this estimator ignores the uncertainty in the predictions $\hat{\bm{x}}_{S}$ used in $\hat{\bm{B}}$, which in turn will underestimate the actual variance of the CNR estimators.
%; we confirm this empirically in Section \ref{sec:simulation}.
% This naive variance estimator can also be used to estimate the variance of the conditional smoother $\beta_0 + \bm{f}_k(x_k)^\top \hat{\bmbeta}_k  + \sum_{l \in \{1,\cdots,K\} \backslash k } \bm{f}_l(\hat{\mu}_l)^\top \hat{\bmbeta}_l $ as
% \begin{equation}
%     \hat{V}_{k,\mathrm{naive}}(x_k) = (1, \bm{f}_1(\hat{\mu}_l)^\top, \cdots, \bm{f}_k(x_k)^\top, \cdots,\bm{f}_K(\hat{\mu}_K)^\top ) \hat{\bm{V}}_{\mathrm{naive}} 
%     \begin{pmatrix}
%         1 \\
%         \bm{f}_1(\hat{\mu}_l) \\
%         \vdots \\
%         \bm{f}_k(x_k) \\
%         \vdots \\
%         \bm{f}_K(\hat{\mu}_K)
%     \end{pmatrix}.
%     \label{eq:V_naive_smoother}
% \end{equation}
% Since the naive variance estimator does not take into consideration the uncertainty and cross-correlation in $\hat{\bm{x}}_S$, it underestimates the variance of the regression coefficient estimators, which 
For $K=1$, \citet{madsenETAL2008} provided a Monte Carlo approach to estimate the variance of their KNR estimator.
% which takes into account the uncertainty of $\hat{\bm{x}}_S$. 
However, their approach is not easily extendable to the more general model \eqref{eq:model_y} involving non-linear associations between the response and multiple spatially cross-correlated covariates as the derivations of their variance estimator largely depend on their assumed simple linear relationship between the response and a single covariate.
% \cite{szpiro2011} study three versions of parametric bootstrap method to correct the variance estimate of the KNR estimator with respect to the measurement error induced by the predicted covariates. \cite{pouliot2023} considers a similar spatial misalignment problem with only one misaligned covariate and proposes an approach based on bootstrap with replacement to conduct inference on the regression coefficient associated with the misaligned covariate, which accounts for the prediction uncertainty of the unobserved covariates due to the estimation of $\hat{\bmtheta}_x$. However, the approach of \cite{pouliot2023} is proposed under the framework of survey sampling which assumes that the variation in the response comes from the random selection of the survey households; that is, the set of locations in $S$. Although such an assumption is appropriate for the dataset of \cite{pouliot2023} which is a survey of Indonesian households, it might not be valid for our motivating dataset whose responses are observed at pollution monitoring stations. 
Instead, we build upon the work of \citet{szpiro2011} and \citet{pouliot2023} and propose a parametric bootstrap method for calculating standard errors 
% (as well as performing bias correction) 
in CNR as follows: first, we perform a preliminary parametric bootstrap to obtain bias-corrected estimates of $\bmtheta_{\rho}$ and $\tau_\epsilon$, which we denote by $\hat{\bmtheta}_{\rho,\mathrm{BC}}$ and $ \hat{\tau}_{\epsilon,\mathrm{BC}}$, respectively. Then, a secondary parametric bootstrap based on these bias-corrected estimates is applied to conduct inference on the regression coefficients. It is important to emphasize this is not a nested or double bootstrap procedure: 
% because the biases of the CNR estimates for the regression coefficients themselves tend to be relatively small, we do not need to employ a computationally intensive double bootstrap, but 
rather it applies two bootstraps sequentially, where the former is solely used to bias-correct the spatial covariance parameter estimates in the spatial linear mixed model; see also Section \ref{sec:appendix_supp_sim} of the supplementary material for further discussion on bias-correcting the CNR estimates for the regression coefficients.

% In both the preliminary and the second bootstrap, we begin by drawing samples of spatially cross-correlated covariates at $S$ and $\tilde{S}$ based on the estimated $\hat{\bmtheta}_x$, along with samples of spatial random effects and residuals at $S$ based on the estimated $\hat{\bmtheta}_\rho$ ($\hat{\bmtheta}_{\rho,\mathrm{BC}}$ in the case of the second bootstrap) and $\hat{\tau}_\epsilon$ ($\hat{\tau}_{\epsilon,\mathrm{BC}}$ in the case of the second bootstrap), respectively. The responses $\bm{y}$ are then constructed based on the bootstrap samples of covariates, spatial random effects, and residuals using model \eqref{eq:model_y} with the regression coefficients set as the CNR estimates $\hat{\bmbeta}$ (note either $\hat{\bmbeta}$  or $\hat{\bmbeta}_{\mathrm{BC}}$ could be used in the case of second bootstrap). Next, we apply CNR procedure on the bootstrapped responses and misaligned covariates, where importantly we treat the bootstrapped covariates at locations $S$ as unobserved. This is then repeated a large number of times.

In the preliminary parametric bootstrap, we draw samples of spatially cross-correlated covariates at locations in $S$ and $\tilde{S}$ based on the estimated $\hat{\bmtheta}_x$, along with samples of spatial random effects and residuals at locations in $S$ based on the estimated $\hat{\bmtheta}_\rho$ and $\hat{\tau}_\epsilon$, respectively. The responses $\bm{y}$ are then constructed based on the bootstrap samples of covariates, spatial random effects, and residuals using equation \eqref{eq:model_y}, where the regression coefficients are set to the CNR estimates $\hat{\bmbeta}$. Next, we apply CNR on the bootstrapped responses and misaligned covariates, where importantly we treat the bootstrapped covariates at locations in $S$ as unobserved. Repeating the above for $T$ times, we let $(\hat{\bmbeta}^{(t)\top}, \hat{\bmtheta}_\rho^{(t)\top}, \hat{\tau}_\epsilon^{(t)})^\top$ denote the resulting bootstrap estimates for the $t$-th sample, where $\hat{\bmtheta}_\rho^{(t)\top} = (\hat{\sigma}_{\rho}^{2(t)}, \hat{\nu}_{\rho}^{(t)}, \hat{\alpha}_{\rho}^{(t)})^\top$ and $t=1,\cdots,T$. 
The bias-corrected estimates of the spatial covariance parameters are then given by 
% $\hat{\bmbeta}_{\mathrm{BC}} = 2 \hat{\bmbeta} - \sum_{t=1}^{T} \hat{\bmbeta}^{(t)} / T$, 
$\hat{\tau}_{\epsilon,\mathrm{BC}} = \exp\{ 2 \log(\hat{\tau}_\epsilon) - \sum_{t=1}^{T} \log(\hat{\tau}_\epsilon^{(t)}) / T  \} $ and $\hat{\bmtheta}_{\rho,\mathrm{BC}} = (\hat{\sigma}^2_{\rho,\mathrm{BC}}, \hat{\nu}_{\rho,\mathrm{BC}}, \hat{\alpha}_{\rho,\mathrm{BC}})^\top$, where $\hat{\sigma}^2_{\rho,\mathrm{BC}} = \exp\{ 2 \log(\hat{\sigma}^2_{\rho}) - \sum_{t=1}^{T} \log(\hat{\sigma}_{\rho}^{2(t)}) / T  \} $, $\hat{\nu}_{\rho,\mathrm{BC}} = \exp\{ 2 \log(\hat{\nu}_{\rho}) - \sum_{t=1}^{T} \log(\hat{\nu}_{\rho}^{(t)}) / T  \} $, and $\hat{\alpha}_{\rho,\mathrm{BC}} = \exp\{ 2 \log(\hat{\alpha}_{\rho}) - \sum_{t=1}^{T} \log(\hat{\alpha}_{\rho}^{(t)}) / T  \} $. Note bias correction is performed at the log scale to ensure the resulting estimates are positive. 

After bootstrap bias-correcting the estimates of spatial covariance parameters, the secondary parametric boostrap procedure is then performed in a similar manner to the preliminary bootstrap described above, with two key differences: 1) we replace $\hat{\bmtheta}_{\rho}$ and $\hat{\tau}_\epsilon$ with $\hat{\bmtheta}_{\rho,\mathrm{BC}}$ and $\hat{\tau}_{\epsilon,\mathrm{BC}}$; 2) if appropriate, we construct the bootstrapped conditional smoothers for the $k$-th covariate as $\hat{\beta}_0^{(t)} + \bm{f}_k(x_k)^\top \hat{\bmbeta}_k^{(t)}  + \sum_{l \ne k } \bm{f}_l(c_l)^\top \hat{\bmbeta}_l^{(t)} $ based on varying $x_k$ for $k=1,\cdots,K$, where $c_l$ is some constant such as the estimated mean of the $l$-th covariate.
After performing the secondary parametric bootstrap, we obtain the empirical quantiles of the bootstrap samples $\{\hat{\bm{\beta}}_{k}^{(t)}: t=1,\cdots,T\}$ for constructing percentile confidence intervals of $\bm{\beta}_{k}$ for $k=1,\cdots,K$, noting for simplicity we have used the superscript $t$ to denote samples from the preliminary and secondary bootstrap procedures. We use the same number of bootstrap samples $T$ in both the preliminary and secondary bootstraps, although this is not essential. 
% where $\hat{\beta}_{lk}^{(t)}$ and $\beta_{lk}$ denote the $l$-th element of $\hat{\bmbeta}_k^{(t)}$ and $\bmbeta_k$, respectively, for $k=1,\cdots,K$. 
Bootstrap percentile confidence bands for the conditional smoother of the $k$-th covariate can be constructed in an analogous manner based on the empirical quantiles of the bootstrap samples $\{\hat{\beta}_0^{(t)} + \bm{f}_k(x_k)^\top \hat{\bmbeta}_k^{(t)}  + \sum_{l \ne k } \bm{f}_l(c_l)^\top \hat{\bmbeta}_l^{(t)}: t=1,\cdots,T\}$  for $k=1,\cdots,K$.  
% from the corrected bootstrap for different values of $x_k$. 
%Note point estimates for $\bm{\beta}_k$ and bootstrapped bias corrections could also be constructed using the bootstrap samples. However here we focus on using the resampling approach solely for uncertainty quantification. 

To summarize, in both the preliminary and secondary parametric bootstraps, simulating the covariates from a multivariate Gaussian distribution with covariance given by \eqref{eq:generalized_kronecker} allows for spatial cross-correlations between covariates in the underlying data generation process. Furthermore, uncertainty from the cokriging predictor of covariates $\hat{\bm{x}}_{S}$ is accounted for by assuming their corresponding bootstrapped values are unobserved and thus each bootstrap dataset is spatially misaligned. We present a formal algorithm for the proposed parametric bootstrap method in Section \ref{sec:appendix_algorithm} of the supplementary material

\section{Numerical Study} \label{sec:simulation}
We performed a simulation study to assess the performance of CNR with the bootstrap approach, comparing it with the commonly used estimator based on nearest-neighbor interpolation, and, in the context of uncertainty quantification, the naive variance estimator, a bootstrap estimator based on the non-bias-corrected CNR estimates of the spatial covariance parameters, and another bootstrap estimator which ignores the cross-correlation between covariates.

For the data generating mechanism, we set the spatial locations in $\tilde{S}$ and $S$ to be the same as the meteorological stations and the pollution monitoring stations, respectively, in the real data application in Section \ref{sec:real_data}, with $M = 243$ meteorological and $N = 796$ pollution monitoring stations. We generated $K = 5$ covariates at locations in both $\tilde{S}$ and $S$ from a multivariate Gaussian distribution with mean parameters $\mu_k = 0$ and the generalized Kronecker product covariance matrix in \eqref{eq:generalized_kronecker}, where $\sigma^2_k = 1$, $\nu_k = 0.5$, $\alpha_k = 0.0015$, $\tau_k = 0.15$ for $k=1,\cdots,5$, and the correlation matrix $\bm{R}$ is specified such that 
\begin{comment}
\begin{equation*}
    \bm{R} = \begin{pmatrix}
        1 & 0 & 0 & 0 & 0 \\
        0 & 1 & 0 & 0 & 0 \\
        0 & 0 & 1 & 0.5 & 0.25 \\
        0 & 0 & 0.5 & 1 & 0.5  \\
        0 & 0 & 0.25 & 0.5 & 1 \\
    \end{pmatrix}.
\end{equation*} 
\end{comment}
the first two covariates were independent of all others while the remaining three covariates had a first-order autoregressive correlation structure with autocorrelation parameter 0.5. 
Given the spatial locations in our real application spanned a large geographic domain, we used the great circle distance in kilometers (km) to measure the separation of locations $\| s - v \|$ in the Mat\'{e}rn correlation function. The choice of $\nu_k = 0.5$ was then made to satisfy the constraint $\nu_k \in (0,0.5]$ to ensure a positive-definite covariance matrix on the sphere \citep{gneiting2013}. Also, we chose $\alpha_k = 0.0015$ to produce moderate spatial correlations i.e., the spatial correlations for pairs of stations that were 438.34km (first quartile of all possible pairwise distances between the $N+M = 1039$ spatial locations) and 1012.15km (third quartile) apart were equal to 0.518 and 0.219, respectively. 

Given the above, we then generated the random effects $\bmrho$ in the spatial linear mixed model from a multivariate Gaussian distribution with zero mean vector $\bm{0}_N$ and covariance parameters $\sigma^2_\rho = 0.2$, $\nu_\rho = 0.5$ and $\alpha_\rho = 0.0015$ for $\bmSigma_\rho$, while the residuals $\bmepsilon$ were generated from a multivariate Gaussian distribution with mean $\bm{0}_N$ and covariance matrix $\tau_\epsilon \bm{I}_N$ with $\tau_\epsilon = 0.01$. The responses were then simulated based on \eqref{eq:model_y}, where for the simulation study we set $\bm{f}_k(x_{ik})^\top \bmbeta_k = x_{ik} \beta_k$ for $k=1,\cdots,5$ and the true regression coefficients as $\bmbeta = (\beta_0,\beta_1,\beta_2,\beta_3,\beta_4,\beta_5)^\top = (2,1,0.5,1,0.5,1)^\top$. A total of 400 simulated datasets were generated. We assumed the smoothness parameters $\nu_\rho$ and $\nu_k; k=1,\cdots,5$ were known.
% function while proving that the quantity $\sigma^2 \alpha^{\nu}$ which is important for interpolation can be estimated consistently when $\nu$ is known. 

Treating the covariate values at locations in $S$ as unobserved i.e., only the response was observed at these locations, we applied CNR to compute the point estimates $\hat{\bmbeta}$ based on the response simulated at locations in $S$ and the covariates generated at spatially misaligned locations in $\tilde{S}$. 
% = (\hat{\beta}_0, \hat{\beta}_1, \hat{\beta}_2,\hat{\beta}_3, \hat{\beta}_4, \hat{\beta}_5)^\top
For the parametric bootstrap procedure in Section \ref{sec:bootstrap}, we set $T=250$ in both the preliminary and secondary parametric bootstraps
% to compute the bias-corrected estimates for the spatial covariance parameters. We then applied the second parametric bootstrap (denoted as Bootstrap, and using $T=250$ also) 
to estimate the standard errors of the CNR estimator $\hat{\bmbeta}$ using the bootstrap standard errors 
and construct bootstrap percentile confidence intervals (CIs) for the regression coefficients based on the 2.5-th and 97.5-th percentiles of the bootstrap samples $\{\hat{\beta}_k^{(t)}: t= 1,\cdots,250\}$ for $k=1,\cdots,5$. 

We compared our point estimates with $L$ nearest-matching-and-regress ($L$-NMR) estimators, which use nearest-neighbor interpolation methods to predict the unobserved covariates at each location in $S$ from the mean of the covariate values observed at the nearest $L$ locations in $\tilde{S}$ with $L \in \{1,3,5\}$ before fitting the spatial linear mixed model \eqref{eq:model_y}. 
We compared our standard errors for $\hat{\bmbeta}$ with those obtained from the naive variance estimator (Naive) $\hat{\bm{V}}_{\mathrm{naive}} = \{\hat{\bm{B}}^\top (\hat{\bmSigma}_\rho + \hat{\tau}_\epsilon \bm{I}_N)^{-1} \hat{\bm{B}}\}^{-1}$ and a bias-corrected version (Naive BC) where $\hat{\bmSigma}_\rho$ and $\hat{\tau}_\epsilon$ were evaluated at the bootstrap bias-corrected estimates of the spatial covariance parameters. A naive variance estimator analogous to $\hat{\bm{V}}_{\mathrm{naive}}$ was similarly used to estimate the standard errors of the $L$-NMR estimators. All of these naive variance estimators ignored the uncertainty in the predicted covariates. Additionally, we performed an unadjusted parametric bootstrap (Unadj-Bootstrap) with $T=250$ to estimate the standard error of $\hat{\bmbeta}$. This approach uses only the secondary bootstrap discussed in Section \ref{sec:bootstrap} i.e., it does not perform a bias-correction on $\hat{\bmtheta}_\rho$ and $\hat{\tau}_\epsilon$. %This was equivalent to skipping the preliminary bootstrap (and bias correction) and going straight into the second bootstrap based on the unadjusted CNR estimates of the spatial covariance parameters, and it was done to investigate the impact of including unadjusted and biased (as discussed in Section \ref{sec:simresults}) spatial covariance parameter estimates in the second parametric bootstrap on the variance estimation of $\hat{\bmbeta}$. 
Finally, we also included a non-cross-correlated parametric bootstrap method (NCC-Bootstrap) with $T=250$, where we replaced $\hat{\bm{R}}$ in $\hat{\bmtheta}_x$ in the secondary parametric bootstrap by $\bm{I}_K$. This method is included to study the effect of ignoring the spatial cross-correlation between covariates on uncertainty quantification in CNR. 
To clarify, for our proposed parametric bootstrap, Unadj-Bootstrap, and NCC-Bootstrap, 95\% bootstrap percentile CIs for the regression coefficients were constructed, while for method Naive, Naive BC, and $L$-NMR, 95\% Wald intervals for the regression coefficients were constructed using their corresponding point estimator and standard error estimator.

 For the regression coefficient of each covariate, we assessed point estimation performance by computing the empirical bias, $\mathrm{Bias}(\hat{\beta}_k) = (1/400)\sum_{l=1}^{400} \hat{\beta}_k^{[l]} - \beta_k $, and empirical root mean squared error, $\mathrm{RMSE}(\hat{\beta}_k) = \sqrt{\mathrm{Bias}(\hat{\beta}_k)^2 + \mathrm{V}_{emp}(\hat{\beta}_k)}$, where $\hat{\beta}_k^{[l]}$ generically denotes an estimator of the $k$-th slope from the $l$-th simulated dataset for $k=1,\cdots,5$ and $l=1,\cdots, 400$, and $\mathrm{V}_{emp}(\hat{\beta}_k) = (1/399)\sum_{l=1}^{400}\{\hat{\beta}_k^{[l]} - (\sum_{l=1}^{400} \hat{\beta}_k^{[l]}/400) \}^2$ denotes the empirical variance of the estimator. For uncertainty quantification, we assessed performance by considering the ratio of average estimated standard error (ASE) to empirical standard deviation (ESD), where 
% ASE is the average estimated standard error using the methods discussed above and 
ESD is given by $\sqrt{\mathrm{V}_{emp}(\hat{\beta}_k)}$. Values of ASE/ESD smaller (larger) than one imply that the estimated standard error is smaller (larger) than the true standard error. Finally, for inferential performance, we examined the empirical coverage probability of the various 95\% CIs for the regression coefficients provided above.

\subsection{Results} \label{sec:simresults}
% As the results for the BC-CNR estimator were highly similar to that of the CNR estimator 
As the 5-NMR estimator was found to perform best among three choices of the $L$-NMR method, then for brevity we only report this estimator below. Results for the 1-NMR and 3-NMR estimators can be found in the supplementary material. For point estimation, Table \ref{table:beta} shows that CNR consistently outperformed the 5-NMR estimator in recovering the relationship between the response and the covariates, with the latter having substantially greater bias and RMSE than the former. 
% In addition, the 1-NMR and 3-NMR estimators in Table \ref{table:beta_full} of Appendix \ref{sec:appendix_supp_sim} suffered from even larger biases.

Turning to uncertainty quantification, all three versions of the naive standard error estimators (two versions for CNR based on whether the unadjusted or bias-corrected estimates of the spatial covariance parameters were used in $\hat{V}_{kk,\mathrm{naive}}$, along with one version for 5-NMR) substantially underestimated the empirical standard error. This is as expected given all of these estimators ignore 
% the cross-correlation between covariates and 
the prediction uncertainty of the unobserved covariates. 
The unadjusted bootstrap standard errors overestimated the empirical standard deviation due to the positive biases in the spatial covariance parameter estimates; indeed in Section \ref{sec:appendix_supp_sim} of the supplementary material
% , which presents the estimation performance for the spatial covariance parameters in the spatial linear mixed model, the 
we present empirical results that demonstrate the positive biases in the CNR (and $L$-NMR) estimators of the spatial covariance parameters, and confirm the effectiveness of preliminary parametric bootstrap in bias-correcting these estimators. 
The proposed parametric bootstrap for CNR gave the most reasonable standard error estimates, supporting the need for bias-correcting the estimates of the spatial covariance parameters and accounting for both the cross-correlation between covariates and prediction uncertainty of $\hat{\bm{x}}_S$. On the other hand, the standard error estimates from the non-cross-correlated bootstrap  
% were reliable for the CNR estimators of the first two covariates (which were independent of the other covariates), but 
underestimated the empirical standard deviation of the slope estimators for the last three covariates (which were all correlated with each other in the true model).

Table \ref{table:CI_coverage} demonstrates that the 95\% CIs for the regression coefficients constructed from CNR and the proposed parametric bootstrap procedure performed best in terms of empirical coverage. There is noticeable undercoverage for the CIs constructed using three naive variance estimators, while
% (relevant to both the CNR and 5-NMR approaches). 
% , and this problem was even more noticeable for the CIs based on the 1-NMR and 3-NMR point estimators due to their larger biases (see Table \ref{table:CI_coverage_full} in Appendix \ref{sec:appendix_supp_sim}). 
% On the other hand, the 
the CIs constructed from the unadjusted bootstrap procedure for CNR tended to be wider than those of the proposed parametric bootstrap leading to overcoverage issues.
The CIs constructed from the non-cross-correlated bootstrap for CNR had empirical coverage close to the nominal level for the regression coefficients of the first two covariates which were both truly independent of all others, but exhibited undercoverage for the coefficients of the remaining three covariates which were truly correlated with each other.
Note also although both the proposed bootstrap and the non-cross-correlated bootstrap CIs provided satisfactory empirical coverage for the first two covariates' slopes, the latter were on average narrower; this is consistent with the latter (correctly) assuming the independence of the first two covariates from the others. 

In summary, our numerical results demonstrated strong estimation performance by the CNR estimators over the $L$-NMR estimators for the regression coefficients, and the importance of bias-correcting the spatial covariance parameter estimates as well as accounting for both the additional variability due to the prediction of unobserved covariates and the potential cross-correlation between covariates when it comes to statistical inference for spatially misaligned data. 
% , as the former properly accounts for the cross-correlation and the uncertainty of the predictions which are ignored by the latter. 

\section{Application to China Air Pollution and Meteorological Data} \label{sec:real_data}
We applied CNR to spatially misaligned air pollution and meteorological data collected across northern and eastern China, with the aim of studying the association between $\text{PM}_{2.5}$ and different climate predictors. $\text{PM}_{2.5}$ represents fine particulate matter with an atmospheric diameter of less than $2.5 \mu$g. Given the well-documented adverse effects of $\text{PM}_{2.5}$ on human health,
% . It has been shown that exposure to $\text{PM}_{2.5}$ can result in increased blood pressure, cognitive deficit, and respiratory and cardiovascular diseases 
% \citep[e.g.,][]{ailshireANDcrimmins2014,liangETAL2014}. 
%popeETAL2002,
there is considerable and continuing interest in understanding the relationship between meteorological conditions and air pollutant concentration 
% is of key importance and remains an active area of research in spatial statistics and epidemiology 
\citep[e.g.,][]{jhunETAL2015,yangETAL2021}. %reichETAL2011 
%zhangETAL2017, borgeETAL2019, 

\begin{figure}[tb]
    \centering
    \caption{Map of China with geographic locations of pollution monitoring stations (in red) and meteorological stations (in blue).}
        \includegraphics[width=0.8\textwidth]{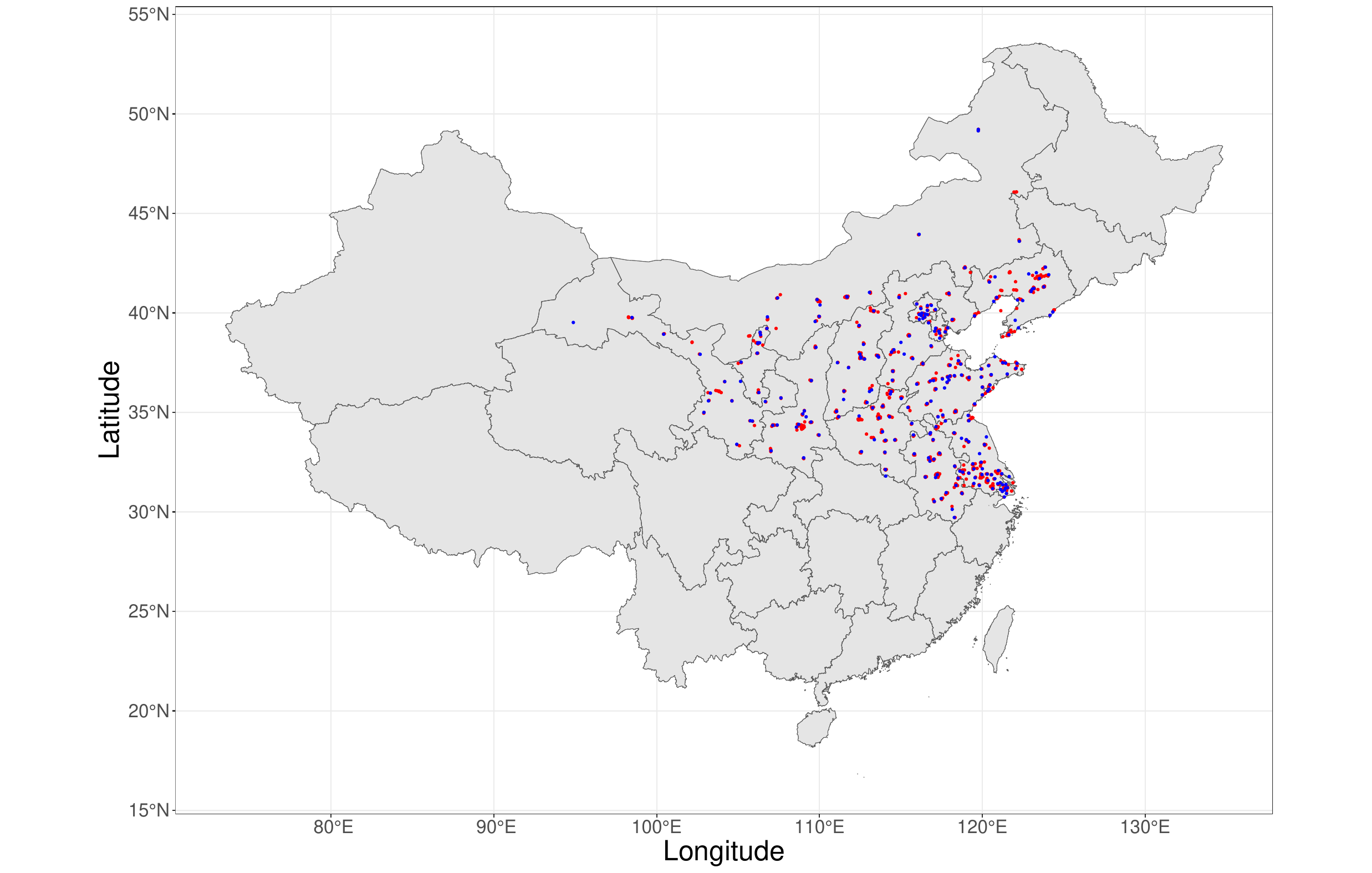}
	\label{fig:pollution_vs_meteorological_locations}
\end{figure}

% According to the 2021 Report on the State of the Ecology and Environment in China (\url{https://www.mee.gov.cn/hjzl/sthjzk/zghjzkgb/}), 35.7\% of cities failed to attain the national air quality standard in 2021, with $\text{PM}_{2.5}$ accounting for the highest proportion (39.7\%) of non-attainment days. It is therefore of interest to study the impact of meteorological covariates on $\text{PM}_{2.5}$ concentration for proper air pollution control. 
The dataset we analyzed were derived from two different sources. The first source provided hourly $\text{PM}_{2.5}$ concentration measurements from $N = 796$ pollution monitoring stations across northern and eastern China.  As the response, we considered the log-transformed mean concentration across the period of July 1 - July 31 2021. 
% We take the log of the averaged $\text{PM}_{2.5}$ concentration (denoted as log transformed PM$_{2.5}$) as the response to improve normality. 
The second source provided weather variables from $M = 243$ meteorological stations at different locations from the pollution monitoring stations (hence the spatial misalignment). These included hourly temperature, precipitation, dewpoint, and wind information in the same period as the $\text{PM}_{2.5}$ concentration. For wind information, based on the original wind data consisting of hourly wind speed and dominant wind direction measured in degrees, we grouped them into four different directions and constructed corresponding variables to measure the average wind speed in each of the four directions; that is, north-west (NW), north-east (NE), south-east (SE), and south-west (SW). For instance, the NW covariate for the $i$-th meteorological station is computed as $\sum_{t=1}^{\mathcal{T}} WS_{it} 1_{\{WD_{it} \in [270 \degree,360 \degree]\}} / \sum_{t=1}^\mathcal{T} 1_{\{WD_{it} \in [270 \degree,360 \degree]\}}$ where $WS_{it}$ and $WD_{it}$ denote the wind speed and the dominant wind direction, respectively, of the $i$-th station at the $t$-th hour for $i=1,\cdots,243$, $t = 1,\cdots,\mathcal{T}$ and $\mathcal{T} = 24 \text{ hours} \times 31 \text{ days} = 744 \text{ hours}$. We then used the average hourly temperature, precipitation, and dewpoint together with these four wind variables as covariates in our model, resulting in $K = 7$ meteorological covariates. Figure \ref{fig:pollution_vs_meteorological_locations} presents the spatial misalignment between PM$_{2.5}$ observed at the pollution monitoring stations, and weather covariates observed at the meteorological stations. 
% Note similar spatial misalignment problems also occurred in the studies mentioned above, and there they simply predicted the unobserved meteorological covariates at the pollution monitoring stations using kriging \citep{reichETAL2011,liuETAL2020}, nearest meteorological station \citep[e.g.,][]{jhunETAL2015,wanETAL2021}, 
%zhangETAL2017,borgeETAL2019,
%or meteorological stations from the same city \citep{liangETAL2016,yangETAL2021}. However, such approaches did not consider the potential spatial cross-correlations between covariates and the prediction uncertainty due to correcting for the spatial misalignment.
% We aim to perform statistical inference on the pollutant-meteorological associations using the CNR method, accounting for spatial misalignment in the presence of multiple correlated covariates.

\begin{comment}
\begin{figure}[tb]
    \centering
        \includegraphics[width=0.8\textwidth]{figure_manuscript/pollution_vs_meteorological_locations.pdf}
	\caption{Map of China with geographic locations of pollution monitoring stations (in red) and meteorological stations (in blue).}
	\label{fig:pollution_vs_meteorological_locations}
\end{figure}
\end{comment}

Based on exploratory analysis (see
% Figure \ref{fig:sample_and_estimated_R}(a) in 
Section \ref{sec:appendix_supp_real_data} of the supplementary material), we found that temperature, precipitation, and dewpoint are moderately correlated, while the remaining four wind variables also presented evidence of moderate cross-correlations between each other. This suggests it is important to account for spatial cross-correlations e.g., through the use of a generalized Kronecker product form in \eqref{eq:generalized_kronecker}. There was also evidence of non-linear relationships between many of the meteorological covariates and PM$_{2.5}$;
% (e.g., Figure \ref{fig:pm25_meteorological_scatter} in Appendix \ref{sec:appendix_supp_real_data}); 
this was consistent with similar non-linear meteorological effects on air pollutant concentration that have been noted in other studies \citep{pearceETAL2011,yangETAL2021}. From these initial findings, we then applied CNR to the spatial linear mixed model \eqref{eq:model_y} with $\nu_\rho = 0.5$ and using natural cubic splines for all covariates, where
%The number of knots for the natural cubic splines was selected based on exploring how sensitive results were for the estimated conditional smoothers; this resulted in 
the knots of the natural cubic splines were placed at the quintiles for each predicted meteorological covariate.
% i.e., for four interior and two boundary knots. 
We also examined a second order polynomial fit with $\bm{f}_k({x_{ik}}) = (x_{ik}, x_{ik}^2)^\top$ for $k=1,\cdots,7$. The results of this fit are provided in the supplementary material, and are broadly consistent with those based on natural cubic splines discussed below.

Table \ref{table:hattheta_x} presents the estimated values of $\hat{\bmtheta}_{x,k}$ obtained in the first step of applying CNR for $k=1,\cdots,7$
% We apply the first two steps of the CNR method in Section \ref{sec:CNR} to estimate $\hat{\bmtheta}_x$ and obtain the prediction $\hat{\bm{x}}_S$ at the pollution monitoring stations based on the observed $\bm{x}_{\tilde{S}}$ at the meteorological stations, 
(fixing $\nu_k = 0.5$ based on exploratory analysis). Dewpoint had a much smaller estimated spatial range parameter than the other covariates, indicating that it had a relatively stronger spatial correlation, while
% compared to the other meteorological covariates. 
temperature and dewpoint had larger estimated variances which reflected the larger scales of these two variables relative to the others.
% as we chose to work with raw meteorological data instead of scaling them. 
The estimated nugget parameters indicated that the four wind variables exhibited much more local variation than the other meteorological variables. The estimated between variable correlation matrix $\hat{\bm{R}}$ is given in 
% (in Figure \ref{fig:sample_and_estimated_R}(b) of 
Section \ref{sec:appendix_supp_real_data} of the supplementary material, and suggests that the generalized Kronecker product form was able to capture the positive correlations among the wind variables as well as the positive correlation between temperature and dewpoint.

\begin{figure}[tb]
\caption{Spatial maps for the cokriging predictor of each meteorological covariate using CNR, based on the observed meteorological data $\bm{x}_{\tilde{S}}$ and the estimated parameters $\hat{\bmtheta}_x$.}
	\includegraphics[width=\textwidth]{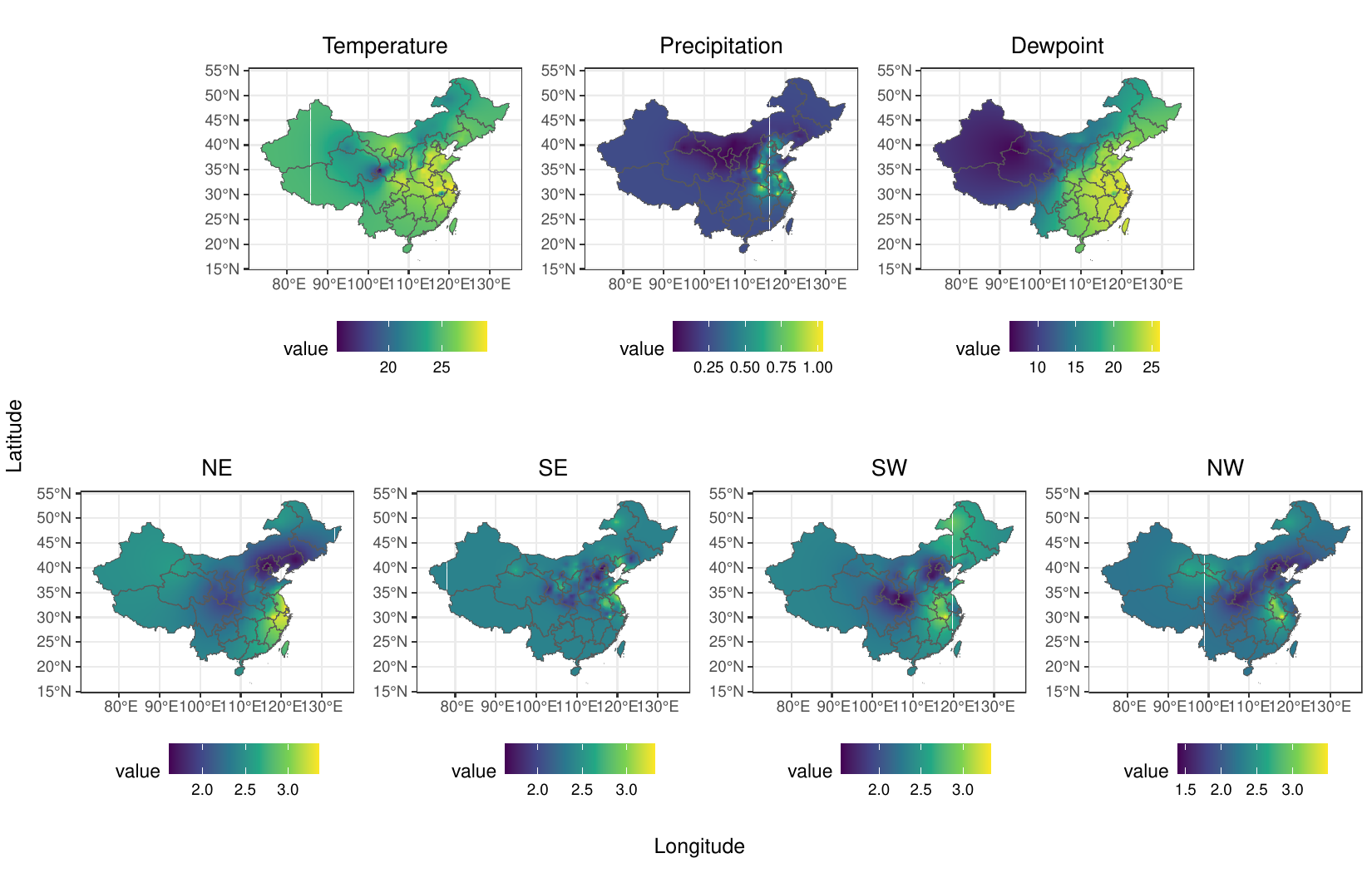}
	\label{fig:prediction_map}
\end{figure}

Figure \ref{fig:prediction_map} presents spatial maps for the cokriging prediction of the meteorological covariates using the observed meteorological data $\bm{x}_{\tilde{S}}$.
% and the estimated parameters $\hat{\bmtheta}_x$. 
These were constructed by discretizing the map of China into a fine grid of pixels each of size $0.1\degree \times 0.1 \degree$ ,
% degree of longitude times 0.1 degree of latitude,
% or equivalently 11 km $\times$ 11 km, 
and then considering the centers of these pixels as the prediction locations. 
The noticeably higher variation in all of the maps towards the eastern part of China can be attributed to the concentration of the observed meteorological stations in that geographical region. Temperature was predicted to be fairly consistent across China, ranging from 15\degree C to 30\degree C in July 2021 during the summer. The map of precipitation showed that eastern China received more precipitation than the other parts; this is related to the flood that occurred in Henan Province in July 2021. In addition, the darker color in the northern part of the precipitation map indicated a much smaller predicted rainfall, which was expected given that this geographic area is part of the Gobi desert. The map of dewpoint demonstrated 
% its stronger spatial correlation compared to the other meteorological covariates due to the smaller spatial range parameter estimate discussed above, and that dewpoint has 
a clear decreasing trend from eastern China to western China. 
% A reasonable explanation for this obvious gradient of predicted dewpoint is that eastern China is surrounded by the Pacific Ocean and the air that moves from the ocean to the land carries more water vapor, i.e., more humid, which results in a higher dewpoint. 
As for the wind variables, eastern China consisting of Shanghai City and Anhui, Jiangsu and Zhejiang provinces had stronger winds due to Typhoon In-fa in July 2021, which also contributed to the aforementioned flood. 
For both the preliminary and secondary bootstrap in Section \ref{sec:bootstrap}, we set $T = 500$ and constructed 95\% percentile confidence bands for the estimated conditional smoothers. 
%Note for the latter, there were two bootstrap datasets which gave extreme 
% for $\hat{\beta}_0^{(t)} + \bm{f}_k(x_k)^\top \hat{\bmbeta}_k^{(t)}  + \sum_{l \in \{1,\cdots,K\} \backslash k } \bm{f}_l(\hat{\mu}_l)^\top \hat{\bmbeta}_l^{(t)} $ 
%estimated spatial range parameters and subsequently extreme conditional smoothers. These were removed when constructing our percentile confidence intervals. 
We refer the readers to the supplementary material for details of the parametric bootstrap procedure when it involves knot placement for splines. For comparison, we used the 5-NMR method to estimate the conditional smoothers of the covariates, and constructed 95\% confidence bands using the naive variance estimators of both CNR and 5-NMR.

\begin{figure}[p]
\caption{Estimated conditional smoothers for the seven meteorological covariates based on CNR (grey solid lines) and 5-NMR (red solid lines) included in the application to the China air pollution and meteorological data. Also shown are 95\% confidence bands based on the naive variance estimator for CNR (shaded regions between dashed lines in grey), bootstrap percentile confidence bands for CNR (shaded regions between dashed lines in yellow) and the naive variance estimator for 5-NMR (shaded regions between dashed lines in red). The top rug represents the observed meteorological covariates $\tilde{\bm{x}}_k$, while the bottom rug represents the predicted meteorological covariates $\hat{\bm{x}}_{k}$ in CNR i.e., based on equation \eqref{eq:cnr_step2_cokriging}.
 % , noting that the training data for conditional smoothers of CNR were $\{\hat{\bm{x}}_k: k = 1,\cdots,7\}$ while the training data for conditional smoothers of corrected parametric bootstrap were $\{\hat{\bm{x}}_k^{(t)}: k = 1,\cdots,7\}$ for $t=1,\cdots,498$.
 }
	\includegraphics[width=0.9\textwidth]{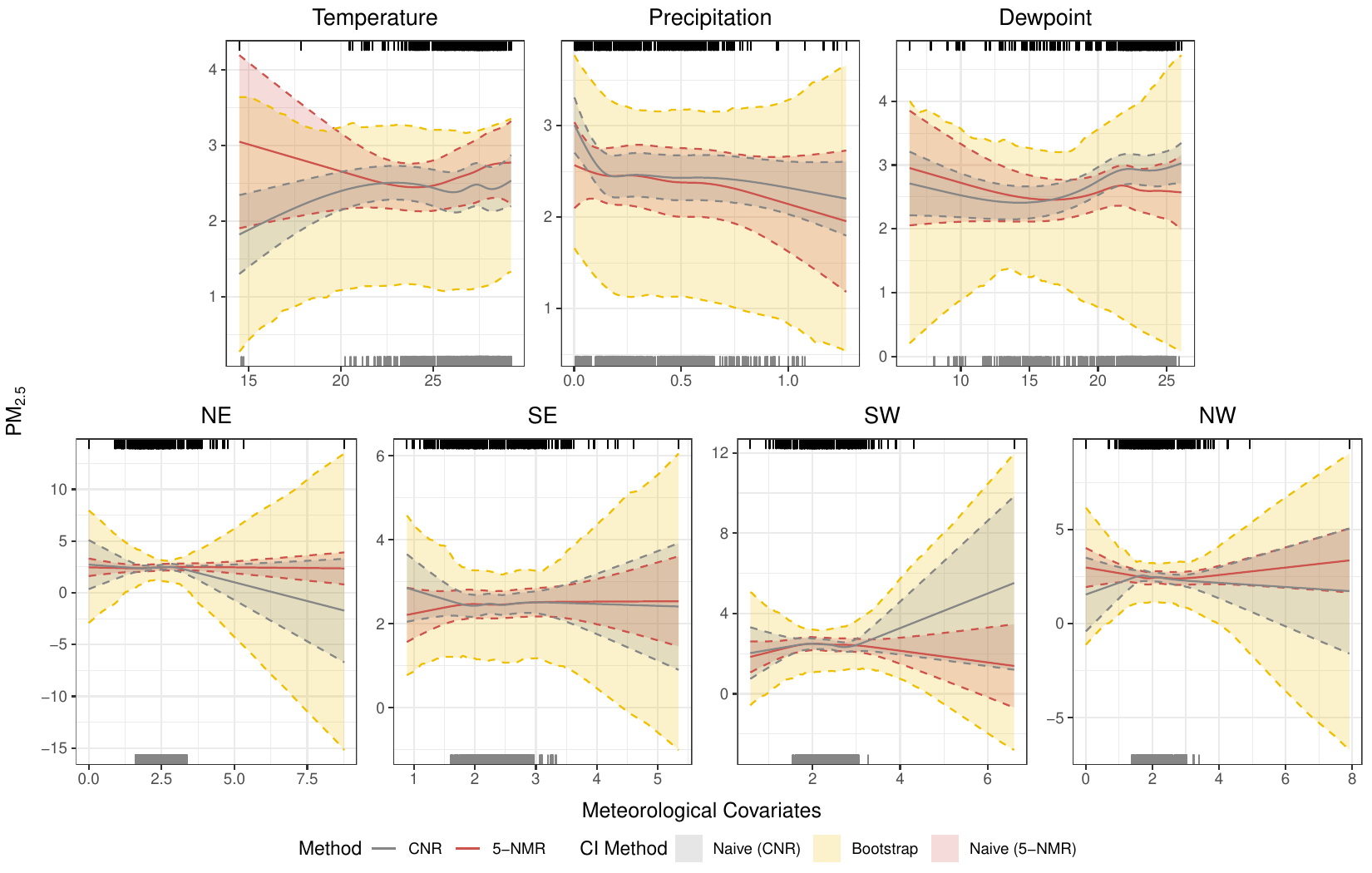}
	
	\label{fig:ns5_results}
\end{figure}

Figure \ref{fig:ns5_results} presents the estimated conditional smoother for each covariate (based on setting the other covariates to their estimated mean values). Although both CNR and 5-NMR methods demonstrated non-linear relationships between the meteorological covariates and $\text{PM}_{2.5}$, their estimated trends were different for some of the meteorological covariates (which may be due to the biases of the 5-NMR estimates seen in the simulation studies of Section \ref{sec:simresults}). %(see also the high similarity between the estimated conditional smoothers of CNR and an additional method - spatial error model in Appendix  \ref{sec:appendix_ns5_real_data}). 
The CNR conditional smoother showed that an increase in temperature in the range of 15\degree C to 20\degree C had a larger positive effect on $\text{PM}_{2.5}$ than the rise in temperature beyond 20\degree C.
% while the 5-NMR conditional smoother demonstrated a negative effect of temperature in the range of 15\degree C to 20\degree C. 
The positive effect is consistent with increases in temperature encouraging photochemical reactions that produce precursors to $\text{PM}_{2.5}$ \citep{jianETAL2012}. As for precipitation, CNR showed an obvious negative impact on $\text{PM}_{2.5}$ consistent with the idea of a washout effect \citep{guoETAL2016}.
% CNR showed a stronger negative effect for lower precipitation levels while 5-NMR estimated an almost linear negative effect. 
Dewpoint exhibited a U-shaped relationship with $\text{PM}_{2.5}$ based on both CNR and 5-NMR.
% , where a similar positive effect has also been noted by \cite{wanETAL2021}. 
Finally, the NE and SW covariates showed negative and positive effects, respectively, based on CNR, while their 5-NMR conditional smoothers suggested that they only have little effect on $\text{PM}_{2.5}$.

\begin{comment}
\begin{figure}[tb]
	\includegraphics[width=0.9\textwidth]{figure_manuscript/ns5_plot1_new_NCC_noSEM.pdf}
	\caption{Estimated conditional smoothers for the seven meteorological covariates based on CNR (grey solid lines) and 5-NMR (red solid lines) included in the application to the China air pollution and meteorological data. Also shown are 95\% confidence bands based on the naive variance estimator for CNR (shaded regions between dashed lines in grey), bootstrap percentile confidence bands for CNR (shaded regions between dashed lines in yellow), and the naive variance estimator for 5-NMR (shaded regions between dashed lines in red),  The top rug represents the observed meteorological covariates $\tilde{\bm{x}}_k$, while the bottom rug represents the predicted meteorological covariates $\hat{\bm{x}}_{k}$ in CNR i.e., based on equation \eqref{eq:cnr_step2_cokriging}.
 % , noting that the training data for conditional smoothers of CNR were $\{\hat{\bm{x}}_k: k = 1,\cdots,7\}$ while the training data for conditional smoothers of corrected parametric bootstrap were $\{\hat{\bm{x}}_k^{(t)}: k = 1,\cdots,7\}$ for $t=1,\cdots,498$.
 }
	\label{fig:ns5_results}
\end{figure}
\end{comment}

Turning to the 95\% confidence bands, 
% we can see that all CIs were narrower in segments with more meteorological data. Importantly, 
the proposed bootstrap confidence bands for CNR were wider than the corresponding bands based on the naive variance estimators of both CNR and 5-NMR for all meteorological covariates. This is consistent with the simulation results in Section \ref{sec:simresults}, which suggested that the proposed bootstrap confidence bands better account for the uncertainty arising from spatial misalignment correction and the cross-correlation between the covariates. 
% from the predicted covariates in the former. 
% The narrower CIs from the naive variance estimator of CNR can be understood as an underestimation of the true variability, as we have empirically shown in Section \ref{sec:simulation}. 
%The CIs constructed from the naive variance estimator of SEM were also narrower than the corrected bootstrap CIs.
% , which was again due to the naive variance estimator ignoring the variability in correcting for spatial misalignment via summarizing the misaligned geostatistical data into county-level areal data. 
The confidence bands based on the naive variance estimator for CNR tended to be wider than the 5-NMR confidence bands near the lower and upper ends of the wind variables; such differences can be attributed to the lack of extreme values in the wind covariates' predictions used for the fitting of the spatial linear mixed model in CNR. 

In the supplementary material, we present additional results for 95\% confidence bands based on the bias-corrected version of the naive variance estimator (Naive BC), the unadjusted bootstrap (Unadj-Bootstrap) and the non-cross-correlated bootstrap (NCC-Bootstrap) procedures. Our results again show that the Unadj-Bootstrap confidence bands were wider than the proposed bootstrap confidence bands, where the latter tended to be wider than the confidence bands based on the Naive BC and NCC-Bootstrap, noting both suffered from undercoverage issues in the simulation study in Section \ref{sec:simresults}.
% , while the independent bootstrap CIs. 
%Finally, we considered an alternative way to overcome the spatial misalignment by
%summarizing the pollution and meteorological geostatistical data into areal data before fitting a spatial error model \citep[SEM,][]{lesageANDpace2009}.  Our results in Appendix \ref{sec:appendix_ns5_real_data} show that the SEM method produced fairly similar estimated conditional smoothers to the CNR method.
% and the 95\% CIs based on the naive variance estimator of SEM were narrower than the bootstrap CIs.

% The results for using second order polynomial terms instead of natural cubic splines of meteorological covariates are provided in Appendix \ref{sec:appendix_quadratic_real_data}. Overall, they again indicate the similarity between the conditional smoothers based on the CNR and SEM methods, along with the narrower CIs constructed from the naive and naive BC variance estimators of CNR, SEM and 1-NMR compared to the bootstrap CIs.
% To summarize, this application illustrates the flexibility of the CNR method and the parametric bootstrap approach in dealing with potentially complex response-covariate associations and conducting proper inferences when the observed response and covariates are spatially misaligned.

\section{Conclusion} \label{sec:conclusion}
We have proposed a cokrig-and-regress method for estimating potentially non-linear relationships in a spatial linear mixed model, where the response and multiple correlated covariates are spatially misaligned. CNR replaces the unobserved covariates with their cokriging predictors before fitting the spatial regression model, where a 
% Mat\'{e}rn class correlation function is used to model the spatial correlation and a 
generalized Kronecker product covariance matrix is employed to account for spatial cross-correlations between the multiple covariates. 
% We estimate the parameters for the resulting joint distribution of all the covariates (across all spatial locations) based on a two-step estimation procedure, after which we compute the cokriging prediction of the unobserved covariates, and then estimate the regression coefficients in the spatial linear mixed model. 
% The CNR estimates for the regression coefficients can be used to construct conditional smoothers to explore the non-linear association between the response and each covariate. 
We also develop a parametric bootstrap approach to both perform bias correction on the CNR estimates of the spatial covariance parameters, and conduct statistical inference on the regression coefficients/smoothing functions. 
% This bootstrap procedure is able to account for the cross-correlation between covariates and the additional variability from the prediction of covariates in the CNR procedure. 
Simulation studies show that CNR performs strongly compared with the commonly used $L$-NMR method in estimating the regression coefficients, 
% with the latter being severely biased. Numerical study also demonstrates the strength of 
while the proposed parametric bootstrap confidence interval offers reasonable empirical coverage probability compared to intervals based on the naive variance estimator, the unadjusted bootstrap using biased spatial covariance parameter estimates, and a bootstrap that ignores the spatial cross-correlation between covariates. Application of CNR to spatially misaligned air pollution and meteorological data in China reveals complex, non-linear relationships between $\text{PM}_{2.5}$ concentration and several meteorological covariates.
% based on the use of natural cubic splines for the meteorological covariates. 
% This application also demonstrates that the confidence intervals for the conditional smoothers constructed using the naive variance estimator are always narrower compared to the parametric bootstrap approach, suggesting that the naive variance estimator potentially underestimates the true variability in the conditional smoothers.

%In additional simulation results presented in Section \ref{sec:appendix_supp_sim} of the supplementary material, we found that bias-correcting $\hat{\bmbeta}$ made very little difference in terms of estimation performance (and not surprisingly, bias correction comes at the price of increasing estimation variance), and so we choose to only bias-correct $\hat{\bmtheta}_{\rho}$ and $\hat{\tau}_{\epsilon}$ before employing them in the second bootstrap procedure. We also performed additional unreported simulations where we used the bias-corrected CNR estimates of both the regression coefficients and spatial covariance parameters in the second bootstrap procedure and the results are very similar to those presented in Section \ref{sec:simresults}.

% The proposed CNR method can be applied to handle spatial misalignment problems in fields other than environmental science, e.g., explaining the effect of local land use variables on stream water quality in the field of hydrology \citep{herlihyETAL1998} and studying the relationship between early childhood meteorological shocks and adult socioeconomic outcomes \citep{macciniANDyang2009}. 
A logical next step would be to extend CNR to handle non-Gaussian and/or areal responses and covariates e.g., 
% This could be achieved, say, 
through 
% the use of copulas \citep{chen2017copula} or 
latent Gaussian models \citep{schliep2013multilevel} or spatial autoregressive models \citep{thoETAL2023}. %bradley2022joint
% Moreover bias correction and uncertainty quantification remain relatively straightforward 
Also, given our proposed bootstrap approach for uncertainty quantification is relatively generic, then we can also handle other modifications for $\bm{f}_k(\cdot)$ such as penalized splines \citep{ruppertETAL2003} and regression trees \citep{breimanETAL1984}, as well as different approaches to dealing with spatial correlations such as fixed rank kriging \citep[see][in the context of correcting for measurement error]{ning2023double} and various flavors of Gaussian processes \citep{datta2016hierarchical}. The proposed CNR and bootstrap method could also be extended directly to handle spatial data that are partially misaligned, where $S \cap \tilde{S} \neq \emptyset$ and $S \neq \tilde{S}$, as well as misaligned spatio-temporal data
% noting that the response and covariates could potentially be misaligned both spatially and temporally 
\citep[see for example][for dealing with temporal misalignment of multiple time series]{cismondiETAL2013,doanETAL2015}.
%cismondiETAL2011
Finally, while this work has focused on the empirical performance of CNR, it would be interesting to theoretically study the asymptotic biases of the CNR estimates, perhaps leveraging from existing work on asymptotic bias analysis for the spatial linear mixed model with covariate measurement errors of \cite{liETAL2009}, and subsequently develop results for the nominal coverage probability of bootstrap percentile intervals.

%\begin{comment}
\section*{Acknowledgments}
ZYT was supported by an Australian Government Research Training Program scholarship. FKCH was supported by an Australian Research Council fellowship DE200100435. AHW is supported by an Australian Research Council Discovery Project DP230101908. This research was undertaken with the assistance of data resources provided by Professor Song Xi Chen’s research group based at Peking University (\url{https://www.songxichen.com}).
%\end{comment}

\section*{Declarations}
\textbf{Conflict of interest}\quad The authors declare no conflict of interest in this article.

\bibliographystyle{apalike} 
\bibliography{bibliography}

\clearpage

\clearpage

\clearpage

\clearpage

\begin{table}[tb]                     
	\centering           
 \caption{Empirical Bias, empirical RMSE and the ratio ASE/ESD for CNR and 5-NMR estimators of $\beta_k$ in the simulation study. For CNR, the variance estimators include the naive variance estimator (Naive), the naive variance estimator but using bias-corrected estimates of the spatial covariance parameters (Naive BC), the proposed bootstrap procedure (Bootstrap), the unadjusted bootstrap procedure (Unadj-Bootstrap) and the non-cross-correlated bootstrap procedure (NCC-Bootstrap).} \scalebox{1}{
		\begin{tabular}{ll rrrrr }                                                                                                           
			\toprule[1.5pt] 
                & & \multicolumn{5}{c}{Point estimation} \\
			    & Method & $\beta_{1} = 1$ & $\beta_{2} = 0.5$ & $\beta_{3} = 1$ & $\beta_{4} = 0.5$ & $\beta_{5} = 1$  \\
			\cmidrule{3-7}
                \multirow{2}{*}{Bias} & CNR & $-0.0068$ & $0.0021$ & $-0.0070$ & $0.0046$ & $-0.0111$ \\ 
                %Unadj-Bootstrap & $-0.0131$ & $-0.0003$ & $-0.0128$ & $0.0064$ & $-0.0176$ \\ 
                %&BC-CNR & $-0.0005$ & $0.0045$ & $-0.0012$ & $0.0028$ & $-0.0046$ \\ 
                %Bootstrap & $-0.0140$ & $-0.0004$ & $-0.0138$ & $0.0063$ & $-0.0175$ \\ 
                %Ind-Bootstrap & $-0.0135$ & $-0.0021$ & $-0.0129$ & $0.0020$ & $-0.0169$ \\ 
                %&1-NMR & $-0.6586$ & $-0.3276$ & $-0.6609$ & $-0.3161$ & $-0.6625$ \\ 
                %&3-NMR & $-0.2859$ & $-0.1445$ & $-0.2825$ & $-0.1320$ & $-0.2929$ \\ 
                &5-NMR & $-0.1529$ & $-0.0752$ & $-0.1591$ & $-0.0665$ & $-0.1580$ \\\\ 
                
                \multirow{2}{*}{RMSE} & CNR & $0.1733$ & $0.1670$ & $0.2068$ & $0.2137$ & $0.1948$ \\ 
                %Unadj-Bootstrap & $0.1735$ & $0.1667$ & $0.2057$ & $0.2122$ & $0.1942$ \\ 
                %&BC-CNR & $0.1743$ & $0.1683$ & $0.2092$ & $0.2167$ & $0.1969$ \\ 
                %Bootstrap & $0.1720$ & $0.1669$ & $0.2056$ & $0.2127$ & $0.1941$ \\ 
                %Ind-Bootstrap & $0.1725$ & $0.1649$ & $0.2062$ & $0.2122$ & $0.1949$ \\ 
                %&1-NMR & $0.6676$ & $0.3460$ & $0.6730$ & $0.3452$ & $0.6752$ \\ 
                %&3-NMR & $0.3366$ & $0.2285$ & $0.3494$ & $0.2522$ & $0.3557$ \\ 
                &5-NMR & $0.2516$ & $0.2113$ & $0.2705$ & $0.2535$ & $0.2872$ \\\\
                
                & & \multicolumn{5}{c}{ASE/ESD} \\
                Method & Variance estimator & $\beta_{1} = 1$ & $\beta_{2} = 0.5$ & $\beta_{3} = 1$ & $\beta_{4} = 0.5$ & $\beta_{5} = 1$ \\
                \cmidrule{3-7}
                \multirow{5}{*}{CNR} & Naive & 0.7495 & 0.7749 & 0.7236 & 0.7811 & 0.7750 \\ 
                 & Naive BC & 0.5856 & 0.6060 & 0.5648 & 0.6091 & 0.6045 \\  
                 & Bootstrap & 1.0896 & 1.0789 & 1.0300 & 1.0628 & 1.0928 \\ 
                 & Unadj-Bootstrap & 1.1984 & 1.1929 & 1.1244 & 1.1739 & 1.2047 \\
                 & NCC-Bootstrap & 1.0122 & 0.9868 & 0.8399 & 0.7687 & 0.8974 \\\\
                
                %\multirow{5}{*}{BC-CNR} & Naive & 0.7446 & 0.7692 & 0.7148 & 0.7703 & 0.7657 \\ 
                % & Naive BC & 0.5817 & 0.6015 & 0.558 & 0.6007 & 0.5973 \\ 
                % & Bootstrap & 1.0824 & 1.071 & 1.0175 & 1.0481 & 1.0798 \\ 
                % & Unadj-Bootstrap & 1.1904 & 1.1841 & 1.1108 & 1.1576 & 1.1904 \\ 
                % & Ind-Bootstrap & 1.0054 & 0.9795 & 0.8297 & 0.758 & 0.8867 \\\\
                
                %1-NMR & Naive & 0.8995 & 0.8799 & 0.8945 & 0.9102 & 0.8707 \\ 
                %3-NMR & Naive & 0.7935 & 0.7891 & 0.7884 & 0.8409 & 0.805 \\ 
                5-NMR & Naive & 0.8096 & 0.8159 & 0.8571 & 0.8541 & 0.7827 \\ 
			\bottomrule[1.5pt]
	\end{tabular}                                                                                                                                                    }

	\label{table:beta}   
	
    \end{table}

\clearpage

\begin{table}[tb]                                    
	\centering
 	\caption{Empirical coverage probability and average interval widths for various 95\% confidence intervals (CIs) of $\beta_k$ in the simulation study. For CNR, the CI methods include the Wald-type intervals using the naive variance estimator (Naive), the naive variance estimator but using bias-corrected estimates of the spatial covariance parameters (Naive BC),  
    bootstrap percentile intervals (Bootstrap) and modified percentile intervals using the unadjusted bootstrap procedure (Unadj-Bootstrap) and the non-cross-correlated bootstrap procedure (NCC-Bootstrap). } 
 \scalebox{0.9}{
		\begin{tabular}{lll rrrrr }                                                                                                           
			\toprule[1.5pt] 
			& Method & CI method & $\beta_{1} = 1$ & $\beta_{2} = 0.5$ & $\beta_{3} = 1$ & $\beta_{4} = 0.5$ & $\beta_{5} = 1$  \\
			\cmidrule{4-8}
                \multirow{6}{*}{Coverage}& \multirow{4}{*}{CNR} & Naive & $0.8400$ & $0.8650$ & $0.8275$ & $0.8875$ & $0.8750$ \\ 
                & & Naive BC & $0.7150$ & $0.7375$ & $0.7175$ & $0.7625$ & $0.7625$ \\ 
                & & Bootstrap & $0.9525$ & $0.9625$ & $0.9600$ & $0.9600$ & $0.9550$ \\ 
                & & Unadj-Bootstrap & $0.9725$ & $0.9725$ & $0.9675$ & $0.9725$ & $0.9825$ \\ 
                & & NCC-Bootstrap & $0.9425$ & $0.9475$ & $0.8900$ & $0.8850$ & $0.9150$ \\ 
                & 5-NMR & Naive & $0.7950$ & $0.8500$ & $0.8175$ & $0.8750$ & $0.8125$ \\ 
                \hline
                \multirow{6}{*}{Average Width}& \multirow{4}{*}{CNR} & Naive & $0.5088$ & $0.5073$ & $0.5861$ & $0.6543$ & $0.5909$ \\ 
                & &  Naive BC & $0.3975$ & $0.3967$ & $0.4575$ & $0.5102$ & $0.4609$ \\ 
                & & Bootstrap & $0.7326$ & $0.6999$ & $0.8264$ & $0.8846$ & $0.8290$ \\ 
                & & Unadj-Bootstrap & $0.8057$ & $0.7768$ & $0.9042$ & $0.9744$ & $0.9136$ \\ 
                & & NCC-Bootstrap & $0.6819$ & $0.6395$ & $0.6772$ & $0.6411$ & $0.6778$ \\ 
                & 5-NMR & Naive & $0.6342$ & $0.6316$ & $0.7350$ & $0.8189$ & $0.7359$ \\
			\bottomrule[1.5pt]
	\end{tabular}                                                                                                                                                        }

	\label{table:CI_coverage}  
	
\end{table}

\clearpage

\begin{table}[tb]                                        
	\centering             
 \caption{Estimated parameters $\hat{\bmtheta}_{x,k}$ of the marginal distribution for the seven meteorological covariates included in the application to the China air pollution and meteorological data.} 
 \scalebox{1}{
 
		\begin{tabular}{l rrrr }                            
			\toprule[1.5pt] 
			Covariates & $\hat{\mu}_k$& $\hat{\sigma}^2_k$ &  $\hat{\alpha}_k$ & $\hat{\tau}_k$ \\
			\midrule
			Temperature & $24.6902$ & $7.3164$ & $0.0028$ & $0.0179$ \\ 
                Precipitation & $0.2460$ & $0.0564$ & $0.0049$ & $0.0041$ \\ 
                Dewpoint & $16.5943$ & $66.1287$ & $0.0002$ & $0.0651$ \\ 
                NE & $2.4590$ & $0.2889$ & $0.0013$ & $0.5260$ \\ 
                SE & $2.3786$ & $0.2060$ & $0.0098$ & $0.2742$ \\ 
                SW & $2.3631$ & $0.2222$ & $0.0026$ & $0.2614$ \\ 
                NW & $2.1930$ & $0.2638$ & $0.0036$ & $0.3933$ \\
			\bottomrule[1.5pt]
	\end{tabular}                                                                 }         
	\label{table:hattheta_x}  
\end{table}

\end{document}